\documentclass{aa}
\usepackage[varg]{txfonts}

\usepackage{graphicx}

\usepackage{color}

\usepackage{soul}
\usepackage{amsmath,amssymb}

\begin{document}

\title{Occurrence and Statistics of IRIS Bursts}

\author{Lucia Kleint\inst{1} \and Brandon Panos\inst{1}}
\institute{University of Geneva, 7, route de Drize, 1227 Carouge, Switzerland}

  \date{Received xxx; accepted xxx}

\abstract{
Small reconnection events in the lower solar atmosphere can lead to its heating, but whether such heating can propagate into higher atmospheric layers and potentially contribute to coronal heating is an open question. We carry out a large statistical analysis of all IRIS observations from 2013 and 2014. We identified ``IRIS burst'' (IB) spectra via a k-means analysis by classifying and selecting \ion{Si}{iv} spectra with superimposed blend lines on top of bursts, which indicate low atmospheric heating. We found that $\sim 8\%$ of all observations show IBs with about 0.01\% of all recorded IRIS spectra being IB spectra. We found varying blend absorption levels, which may indicate different depths of the reconnection event and heating. IBs are statistically visible with similar properties and timings in the spectral lines \ion{Mg}{ii} , \ion{C}{ii}, and \ion{Si}{iv}, but invisible in \ion{Fe}{xxi}. By statistically analyzing co-spatial AIA lightcurves, we found systematic enhancements in AIA 1600 and AIA 1700, but no clear response to bursts in all other AIA wavelengths (94, 131, 171, 193, 211, 304, 335) in a timeframe of $\pm 6$ minutes around the burst. This may indicate that heating due to IBs is confined within the lower atmosphere and dissipates before reaching temperatures or formation heights covered by the hotter AIA lines. Our developed methods are applicable for statistical analyses of any co-observed data sets and allow us to efficiently analyze millions of spectra and lightcurves simultaneously.
}

\maketitle

\section{Introduction}

Magnetic reconnection is ubiquitous in the solar atmosphere, where it can lead to an energy release on various scales, from flares to small-scale brightenings low in the solar atmosphere. Such events may temporarily heat the surrounding plasma. It is still an open question whether certain types of events, namely nanoflares, contribute to the unsolved coronal heating problem \citep[e.g.,][]{Klimchuk2006}. It is therefore of significant interest to investigate reconnection and heating events in the solar atmosphere statistically.

\citet{peteretal2014science} reported short-lived ($\sim$5 min) bursts (IRIS bursts, IB) that are thought to heat the gas of the normally few thousand Kelvin lower atmosphere to temperatures of nearly 100,000~K. Such bursts were identified through their particular double-peaked spectral shapes, which were interpreted as bi-directional flows in the lower solar atmosphere after a reconnection event, which subsequently expanded the overlying cooler atmosphere, resulting in absorption lines (e.g. \ion{Fe}{ii}, \ion{Ni}{ii}) superimposed on the burst spectra. Their exact height is debated \citep{judge2015}, but they likely occur a few hundred kilometers above the solar surface with recent modeling efforts placing them in the mid to upper chromosphere \citep{hansteenetal2019, hongetal2021}. No coronal emission was identified by \citet{peteretal2014science}, thus restricting the events to the ``lower'' (meaning below the transition region) solar atmosphere, similar to the so-called Ellerman bombs \citep[EBs,][]{ellerman1917,georgoulisetal2002}. Such events are in contrast to so-called "explosive events" that may show similarly broadened and non-Gaussian line shapes, but lack the superimposed absorption lines, and are interpreted as events occurring in the transition region \citep[e.g.][]{huangetal2017}.

Subsequent investigations classified the properties of IBs in more detail \citep[see review by][]{youngetal2018}, such as their relation and differences to EBs \citep{vissersetal2015,tianetal2016}, their dynamics \citep{lucetal2017}, their recurrence, and also found instances where bursts were potentially visible as brightenings in SDO/AIA data \citep{guptatripathi2015}. The latter study found a time-lag of about 5 minutes between the bursts' appearance in IRIS images and brightenings at a similar location in coronal AIA channels ($\sim$2 MK), but was unable to definitively conclude whether the brightenings were related. This relation, if it exists, is of critical importance, and might suggest that IBs contribute to the heating of the upper solar atmosphere.

Several open questions still remain; how frequent are IBs? How do they develop and evolve statistically, and in particular, at which temperatures and heights do they appear? These questions can be investigated by (1)  finding all IB occurrences automatically. (2) Analyzing them with spectral lines that form at different temperatures, and (3) Correlating the IBs to the evolution of coronal lines of SDO/AIA.  Since this research requires the processing of millions of spectra and images, it is vital to have efficient tools to make the analysis tractable. Such a tool-set was recently developed to analyze co-occurring data types \citep{panosetal2018,panosetal2021,panoskleint2021}. We apply these methods to the investigation of IBs and their possible contribution to coronal heating.

 \begin{figure} 
  \centering 
   \includegraphics[width=.5\textwidth,clip,trim={.3cm 0.5cm 0.5cm 0.5cm}]{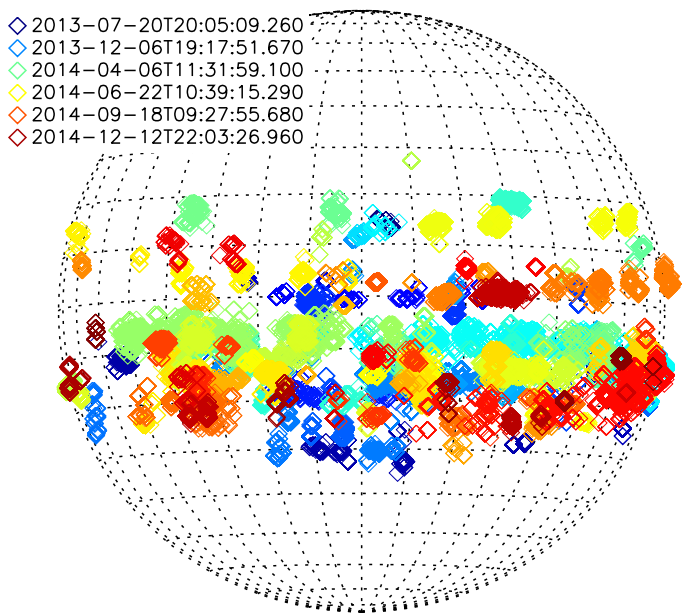}
   \caption{The coordinates of 101337 burst spectra observed by IRIS in 2013 and 2014. The color-coding of the bursts refers to their timestamp, with timestamps for 6 of the bursts shown in the legend.}
        \label{fig1b}
  \end{figure}

\section{Observations and data reduction}
\label{obs}
\subsection{Selection of IRIS burst spectra}
The Interface Region Imaging Spectrograph \citep[IRIS,][]{iris2014} was launched in July 2013 and observes two spectral regions, the FUV (1332 -- 1358 \AA, 1389 -- 1407 \AA) and the NUV (and 2783 -- 2834 \AA), in highly flexible observing modes where exposure times, number of raster steps, repetitions and spectral lines of interest can be customized. The IRIS database contains an archive of various targets and is ideally suited for statistical analyses.

We have analyzed all IRIS observations from 2013 and 2014 in a 3-step procedure to identify IBs with high reliability. The years were selected to obtain a large enough sample of observations when solar activity was high. Step (1): We identified observations with the following criteria: (a) the spectral line \ion{Si}{iv} 1393.75 \AA\ needed to be available to reliably identify bursts based on blend lines in absorption, (b) observations of exposure times $< 0.6$ s, and times within observations when IRIS was passing through the south atlantic anomaly were excluded to avoid noise contamination.
This led to a sample of 3537 observations, many of which spanned several hours. 

Step (2): From these observations, every single \ion{Si}{iv} 1393.75~\AA\ spectrum (a total of $1.16 \cdot 10^9$ spectra) was classified using the nearest neighbors classification\footnote{\url{https://scikit-learn.org/stable/modules/neighbors.html}} and a predefined list of potential spectral shapes to identify candidates for IBs. This classification was performed on normalized spectra to focus on their shape. Only observations with at least 30 promising spectra were kept for further analysis. A total of 287 observations (8.1\% of the analyzed observations) containing 160k spectra of interest remained after this selection step. This first classification was done with minimal restrictions to avoid missing any potentially interesting spectra.

 \begin{figure*} 
  \centering 
   \includegraphics[width=\textwidth]{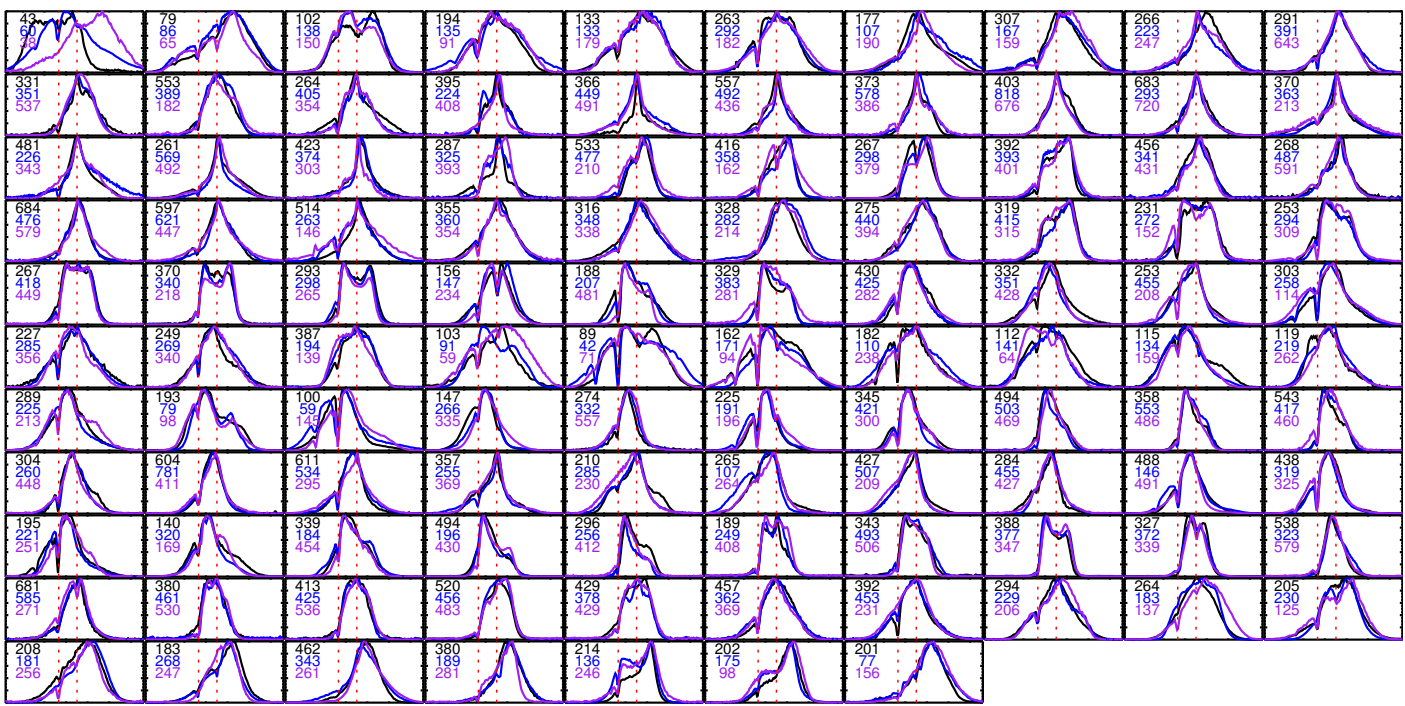}
   \caption{321 \ion{Si}{iv} ``centroids'' (or rather median spectra) depicting the different spectral shapes that were found for IBs. Three spectra are plotted in different colors in each panel to save space, but during the classification all of them were treated independently. The numbers indicate how many spectra were classified into a given centroid. The red dotted lines indicate the wavelengths of line center and of the \ion{Ni}{ii} blend at 1393.33 \AA.}
        \label{fig1}
  \end{figure*}

Step (3):
After interpolating all spectra onto a common wavelength scale, a k-means classification \citep{MacQueen1967} was performed fully automatically with 500 centroids on the 160k spectra of interest.  Analogous to step 2, normalized spectra were used  to focus the classification on line shapes rather than varying intensities. From the 500 obtained centroids, we visually excluded centroids associated with spectra that did not contain absorption blends in \ion{Si}{iv} 1393.75~\AA, a criterion for IBs identified by \citet{peteretal2014science}. We also excluded any overexposed spectrum, defined as containing 5 or more saturated intensity values. 

The final step led to 101337 spectra that very likely are related to bursts, which was also verified visually with random samples. This indicates that about 0.01\% of all IRIS spectra are IB bursts with absorption line blends. Their locations are shown in Fig.~\ref{fig1b} with the color-coding denoting when they were observed (only a few dates are shown in the legend). Their appearance at active latitudes is obvious and also visually, we did not find any events in pure quiet Sun observations. In Figure~\ref{fig1}, we plot the final spectral shapes of bursts, but instead of averaging over all spectra in a given k-means group, as would be the typical definition of a centroid, we show for each k-mean group the spectrum with the minimum distance to the group centroid (often referred to as the medoid). This way, the plot shows 321 real observed spectra that are representative for each burst type that may occur. 
The numbers indicate the distribution of the spectra into the different shapes, showing a somewhat balanced occurrence of the different line shapes. We provide an IDL save file listing the 101337 identified burst spectra and their properties and coordinates\footnote{\url{https://github.com/lkleint/irisbursts}}.

\subsection{Other IRIS wavelengths}
Our goal is to analyze bursts at different heights, temperatures, and their temporal evolution. For each identified \ion{Si}{iv} burst, we therefore extracted the co-occurring spectra from several different IRIS lines. In the FUV, we selected the prominent spectral lines \ion{C}{ii} 1334.53 \AA\ and 1335.71 \AA,  \ion{Si}{iv} 1402.77 \AA, and the coronal \ion{Fe}{xxi} 1354.08 \AA\ line. In the NUV, we included spectral lines of \ion{Mg}{ii}, namely the resonant k line (2796.35 \AA) and the triplet at 2791.60, 2798.75 and 2798.82 \AA, as well as the \ion{Fe}{ii} 2814.445 $\text{\AA}$ line. All spectra were interpolated onto common wavelength scales. For observations where e.g.\,\ion{Fe}{ii} or \ion{Fe}{xxi} was not recorded due to different line lists, these spectra were set to -1. Similarly, overexposed spectra were flagged as -1 and did not participate in any subsequent calculations.


\fboxsep=0.5mm
\fboxrule=0.5pt

 \begin{figure*} 
  \centering 
   \includegraphics[width=.19\textwidth]{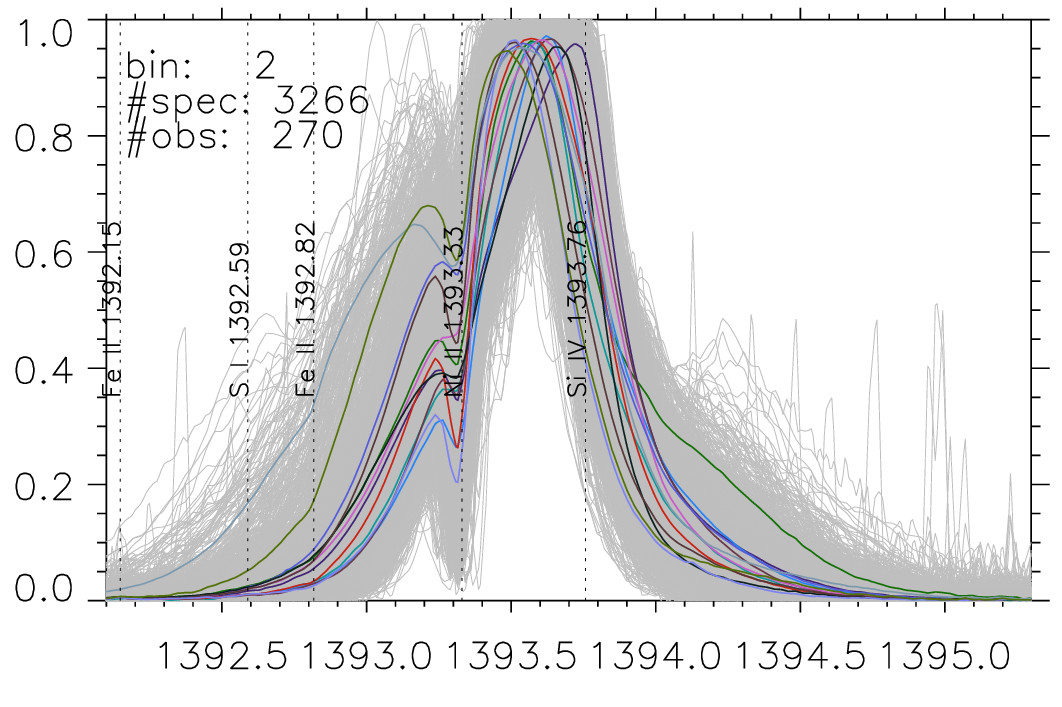}
   \includegraphics[width=.19\textwidth]{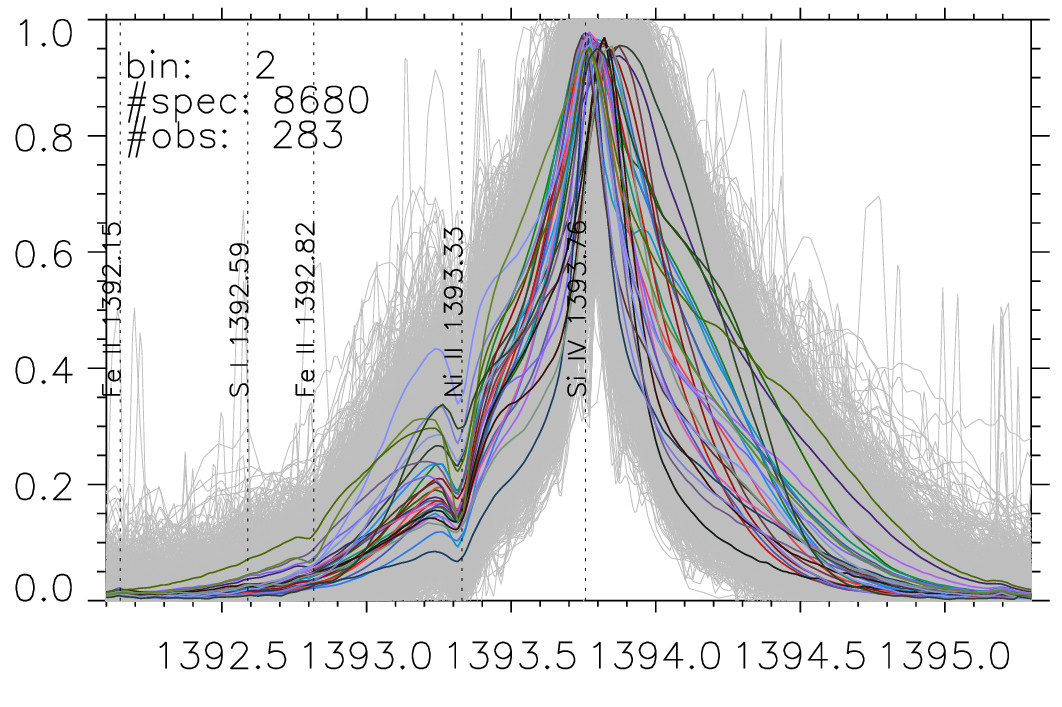}
   \includegraphics[width=.19\textwidth]{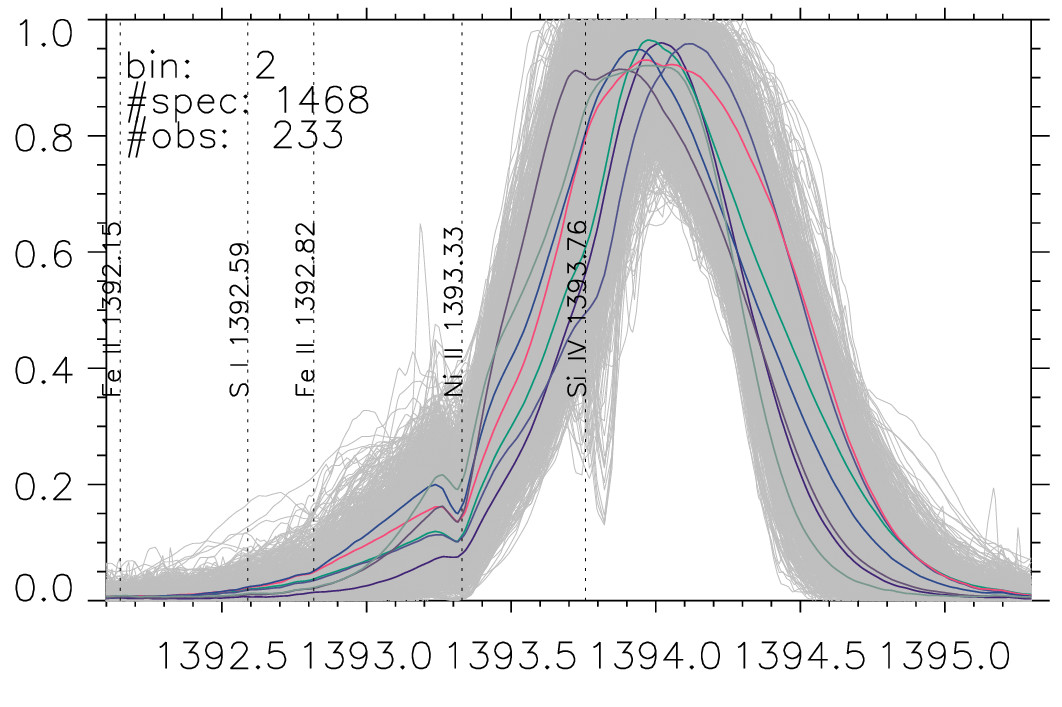}
   \includegraphics[width=.19\textwidth]{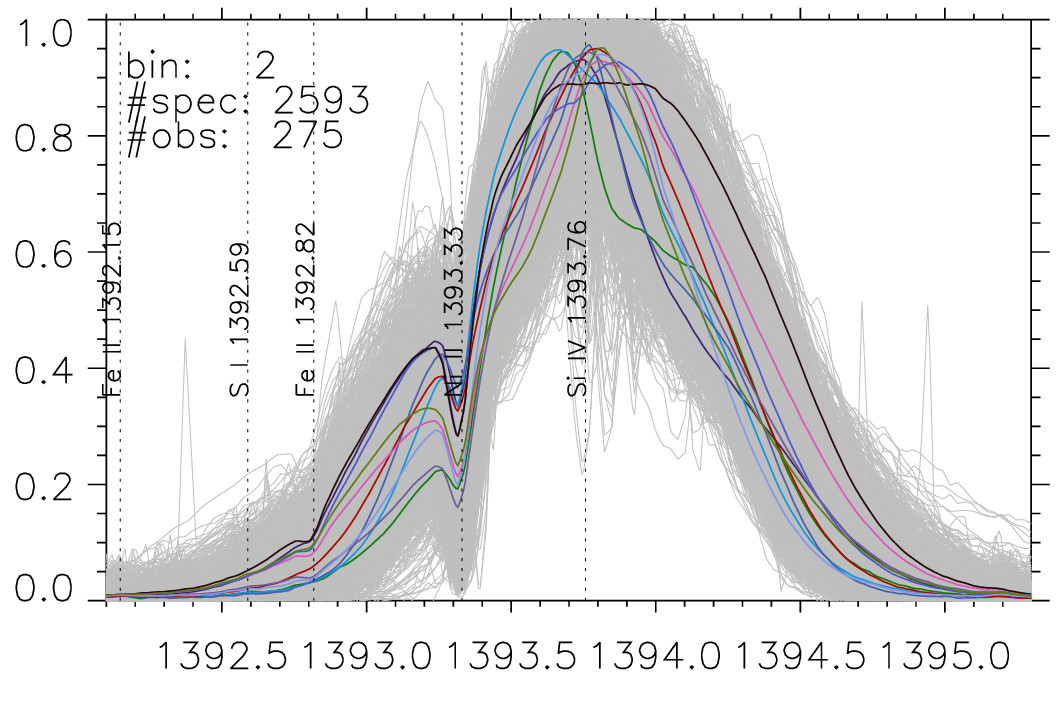}
   \fcolorbox{white}{yellow}{\includegraphics[width=.19\textwidth]{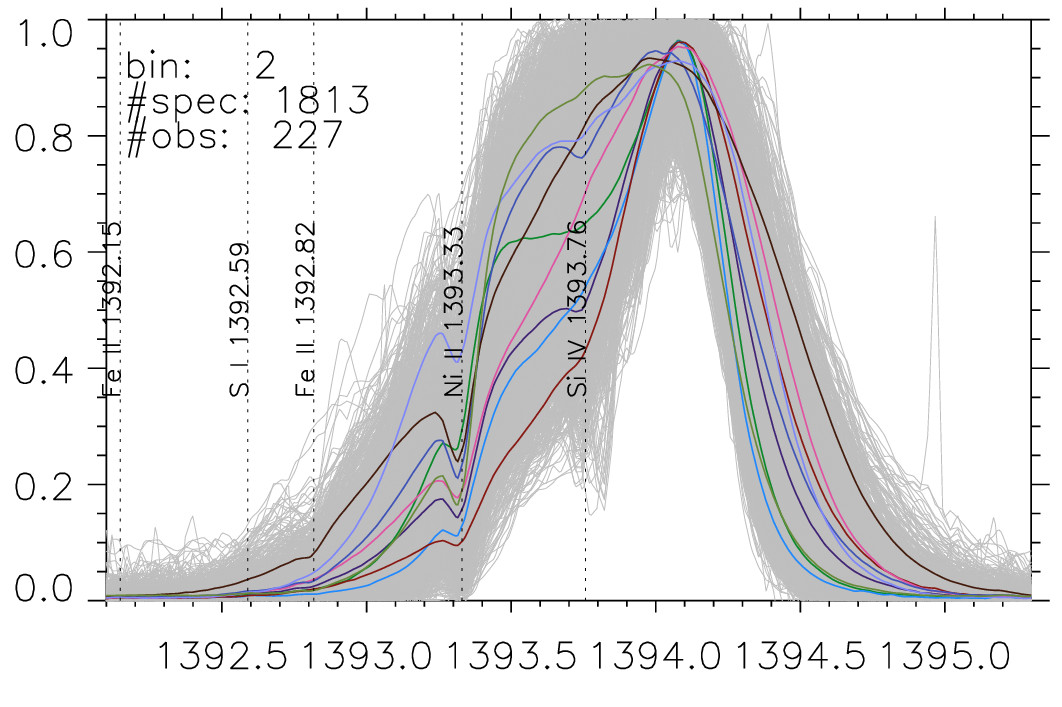}}
   \includegraphics[width=.19\textwidth]{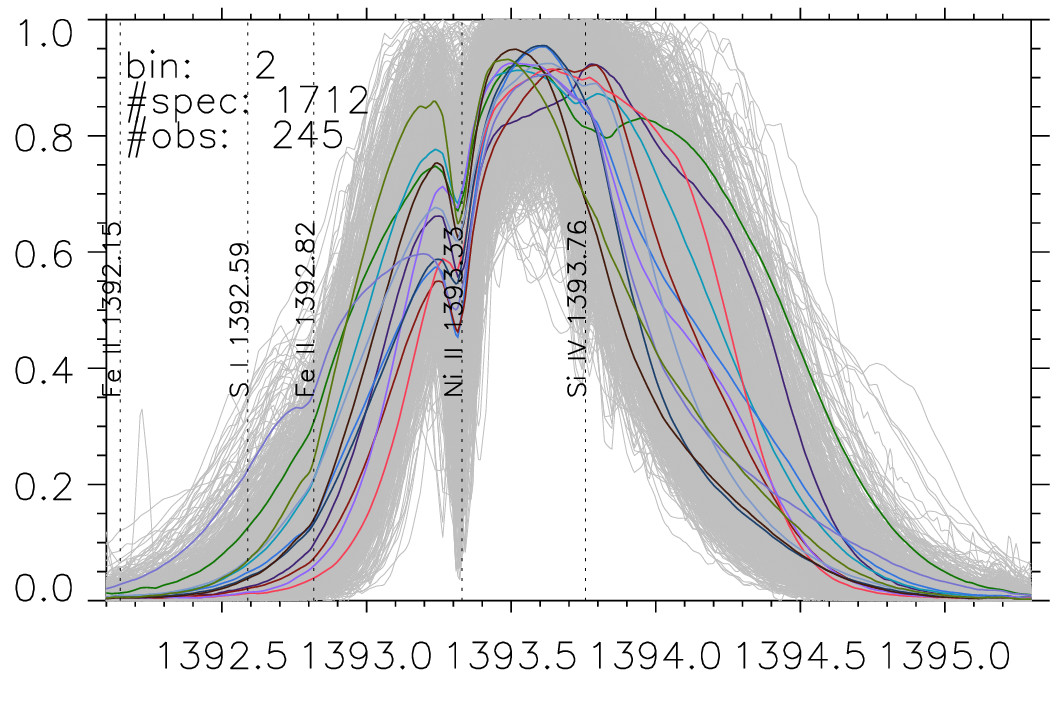}
   \includegraphics[width=.19\textwidth]{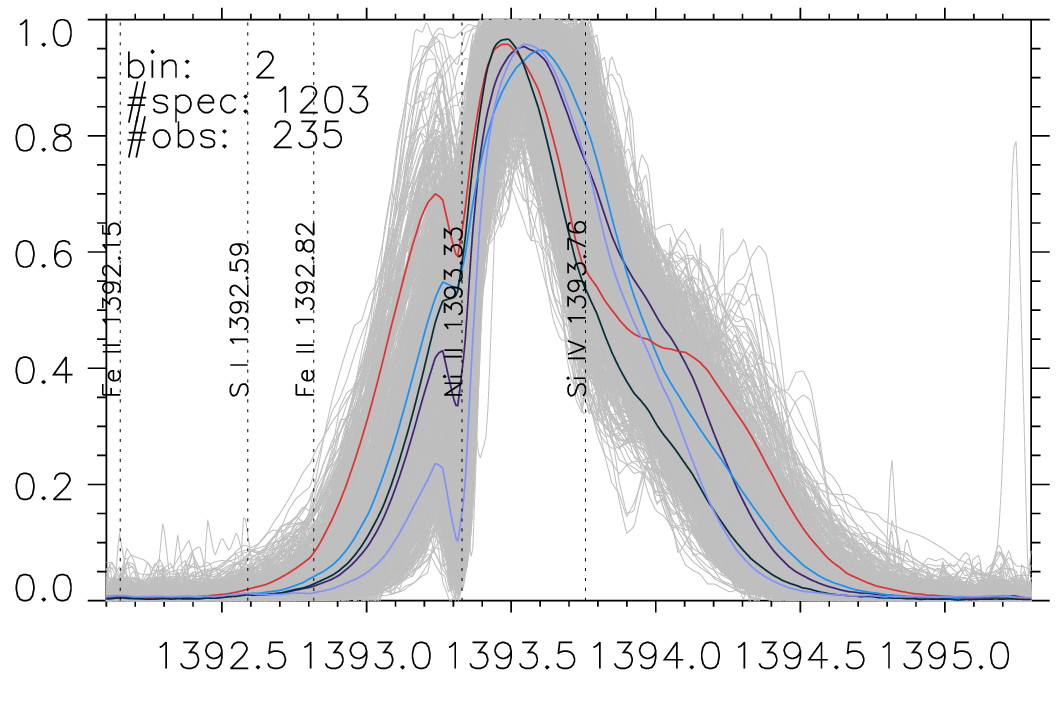}
   \includegraphics[width=.19\textwidth]{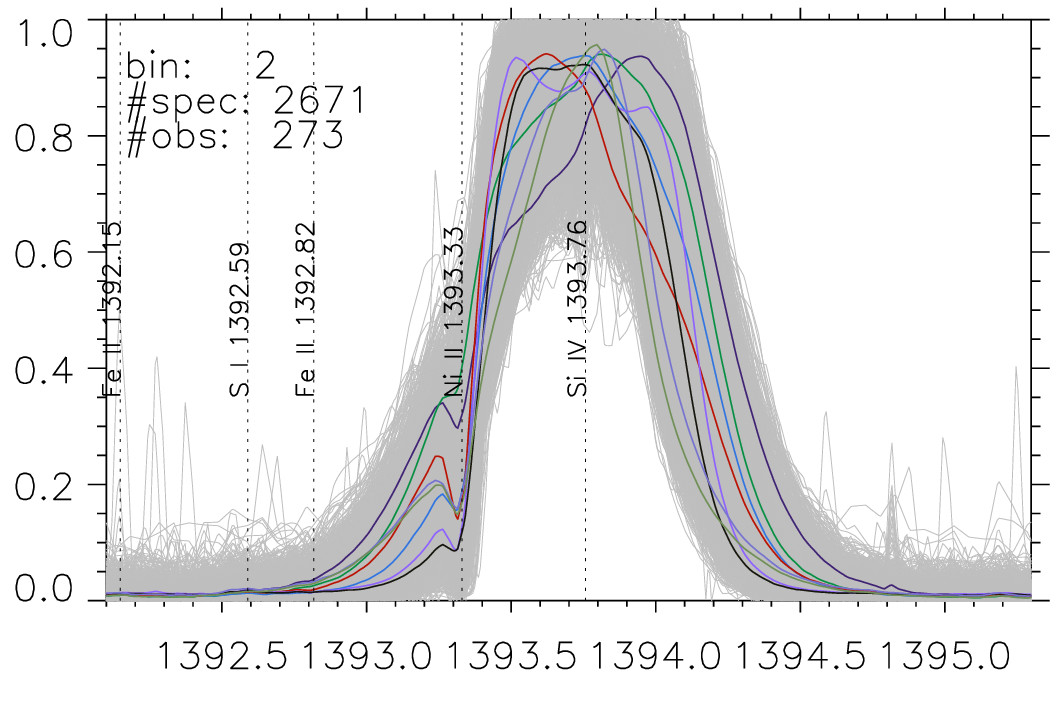}
   \includegraphics[width=.19\textwidth]{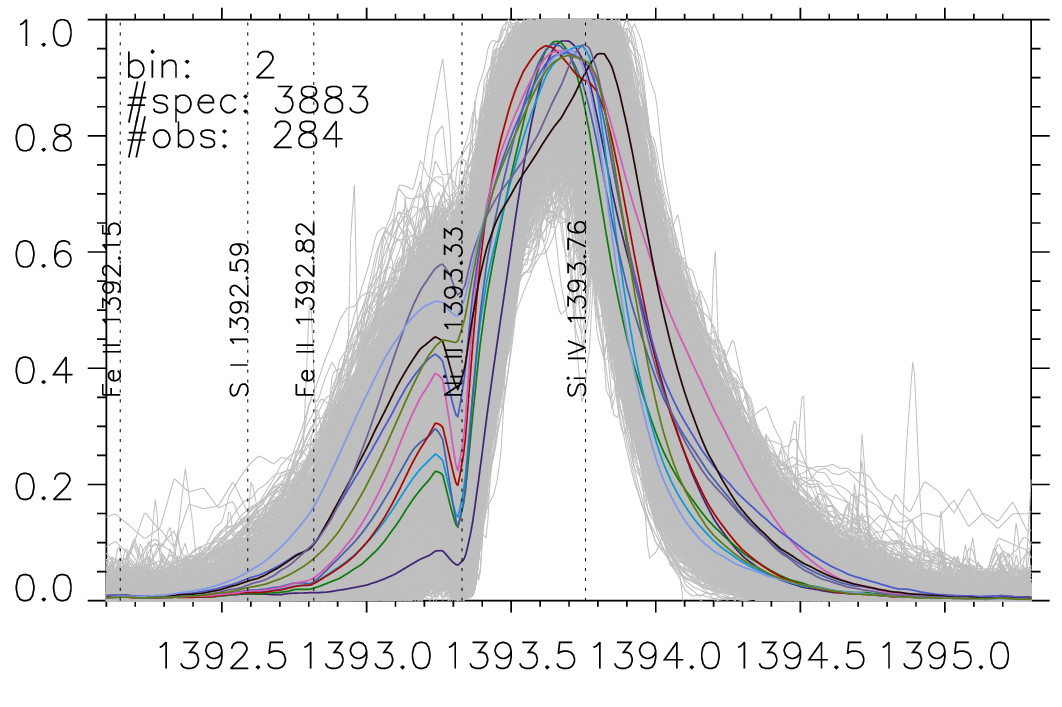}
   \fcolorbox{white}{yellow}{\includegraphics[width=.19\textwidth]{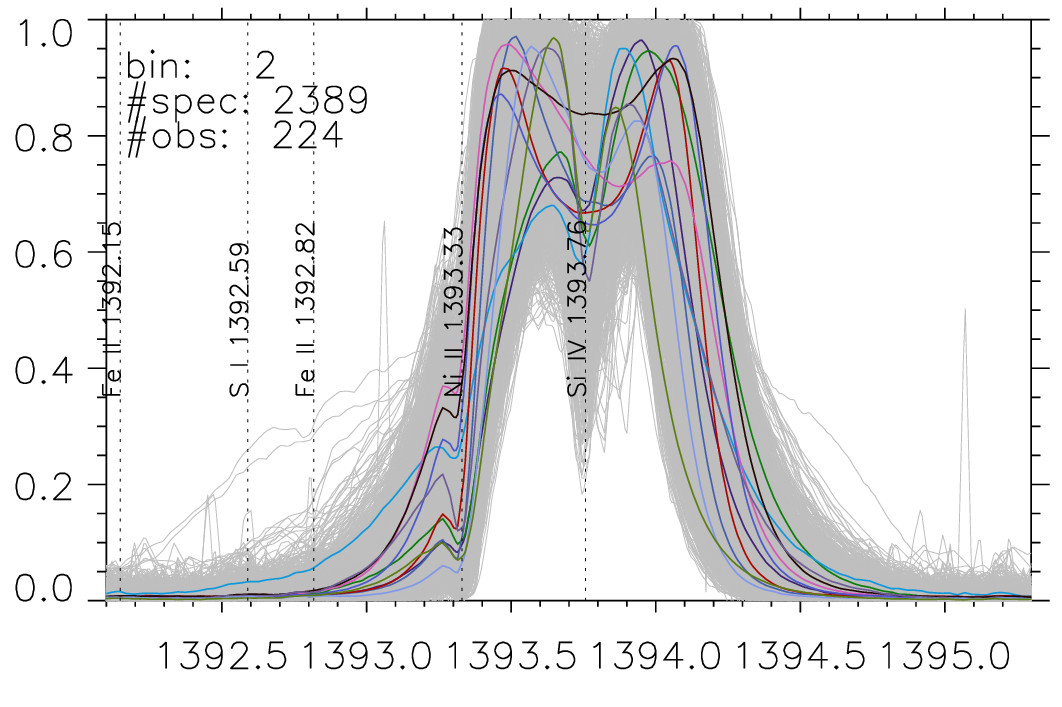}}
   \includegraphics[width=.19\textwidth]{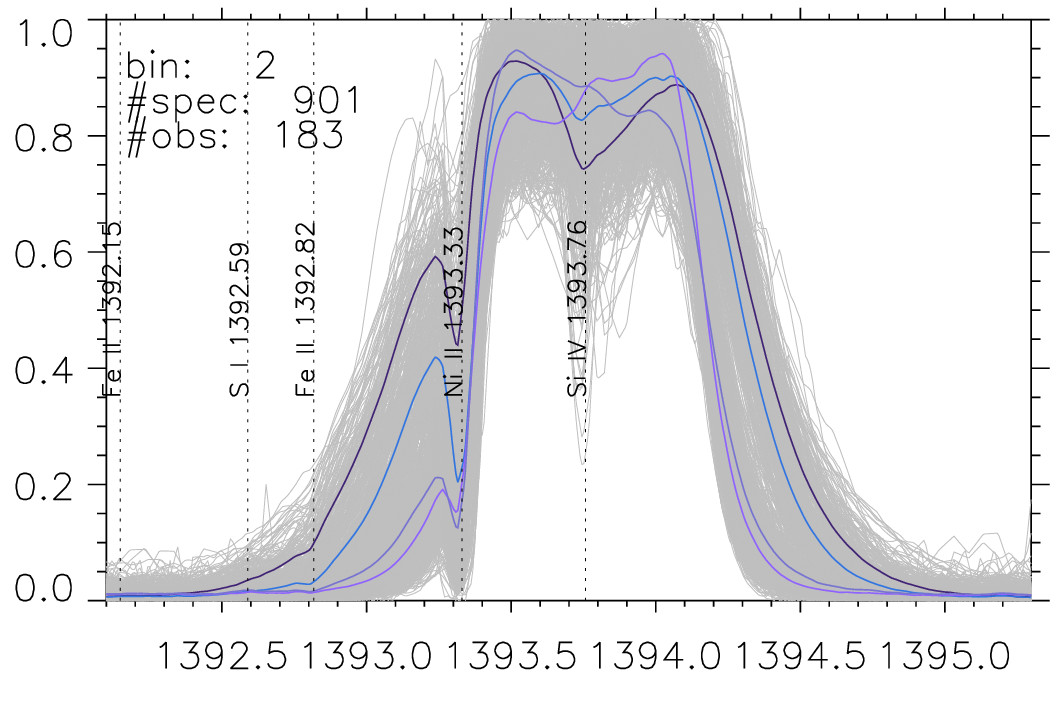}
   \includegraphics[width=.19\textwidth]{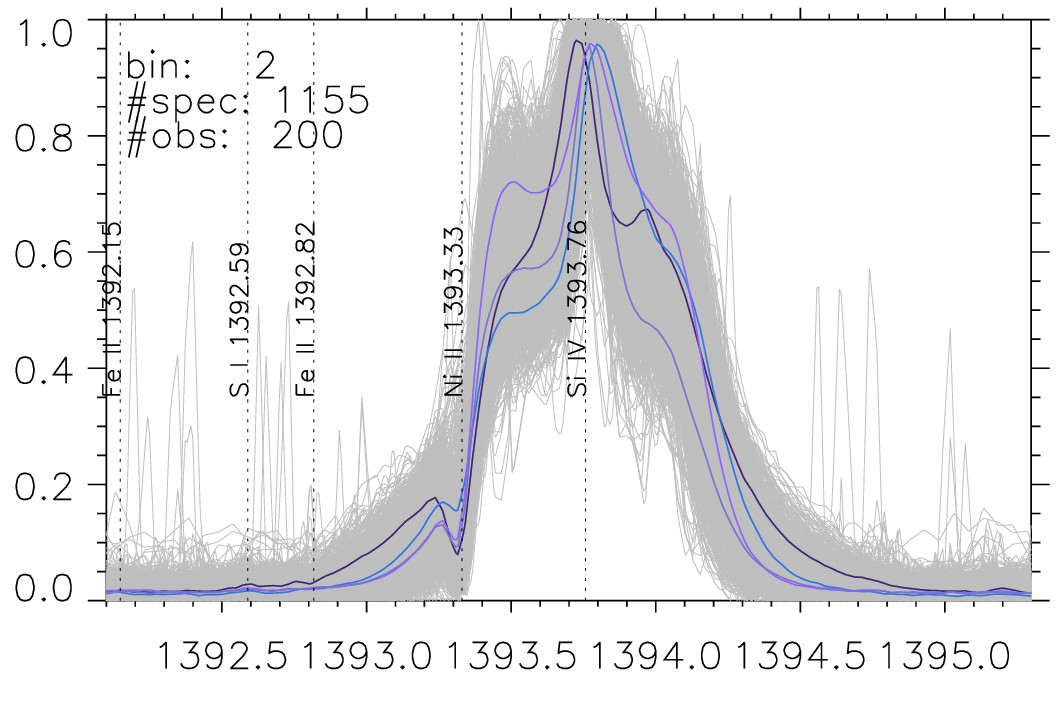}
   \fcolorbox{white}{yellow}{\includegraphics[width=.19\textwidth]{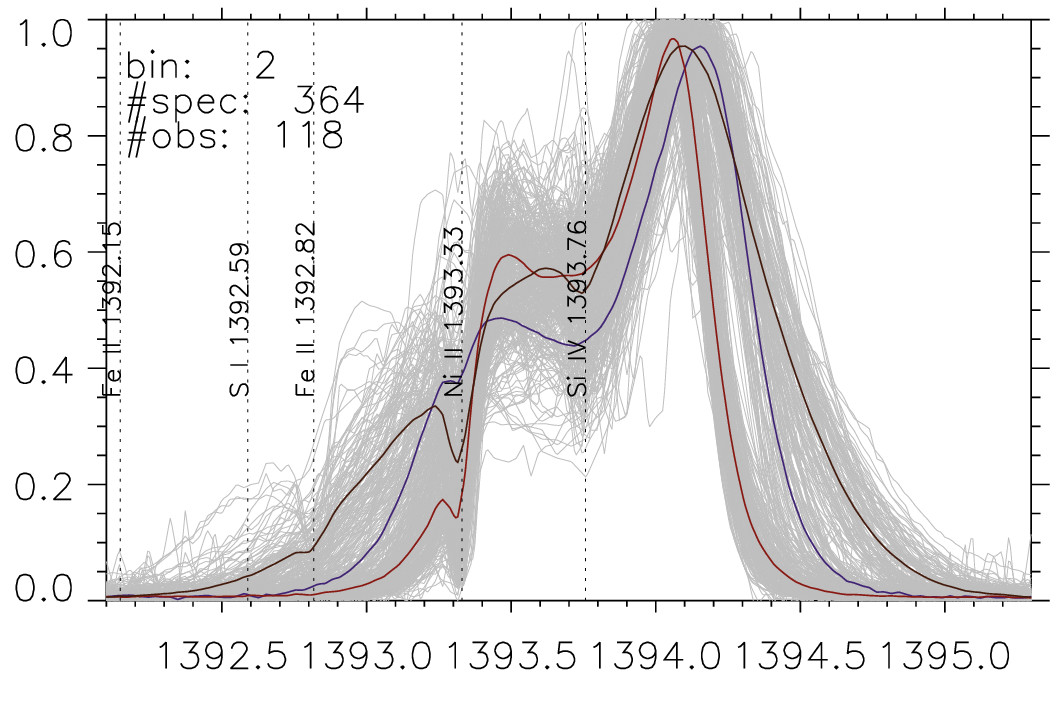}}
   \includegraphics[width=.19\textwidth]{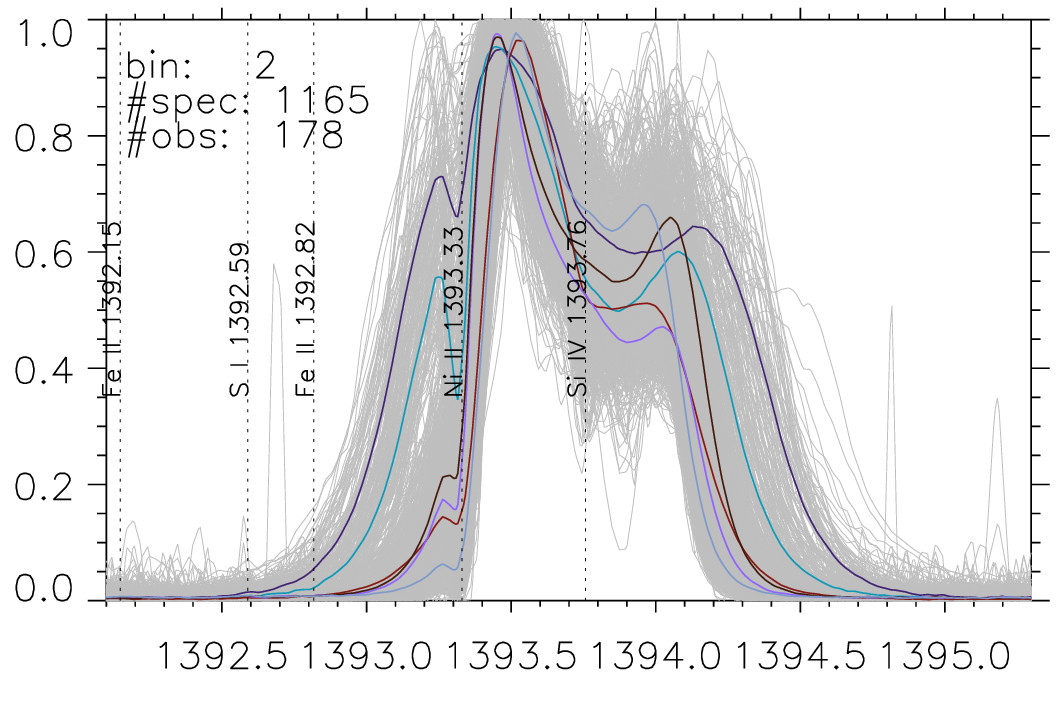}
   \fcolorbox{blue}{yellow}{\includegraphics[width=.19\textwidth]{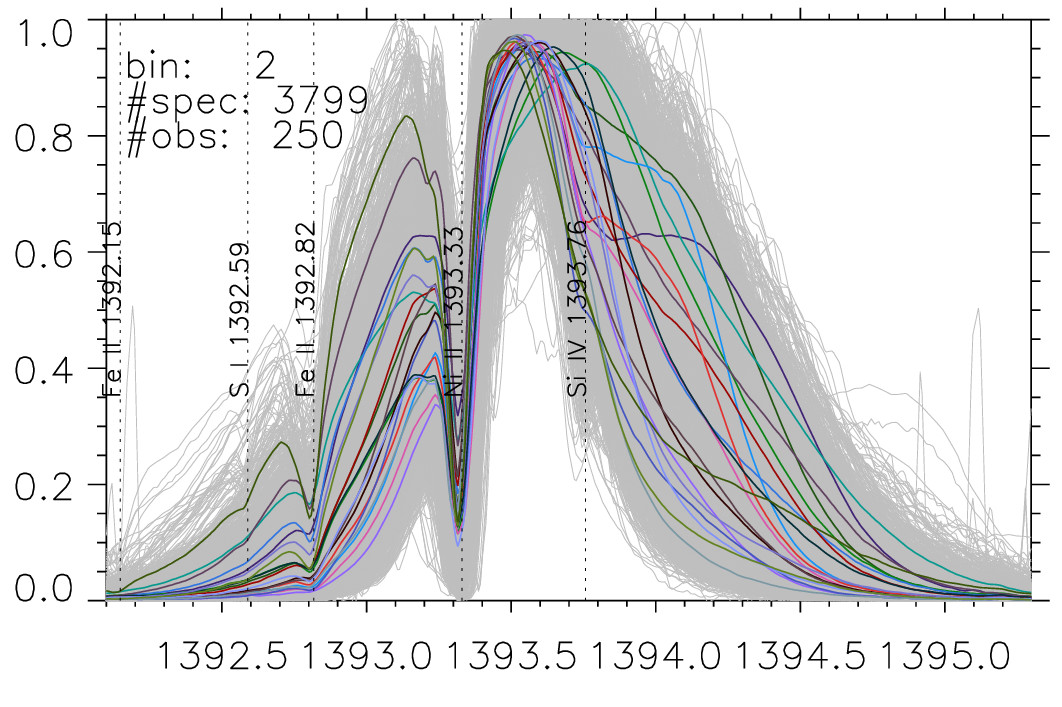}}
    \fcolorbox{blue}{white}{\includegraphics[width=.19\textwidth]{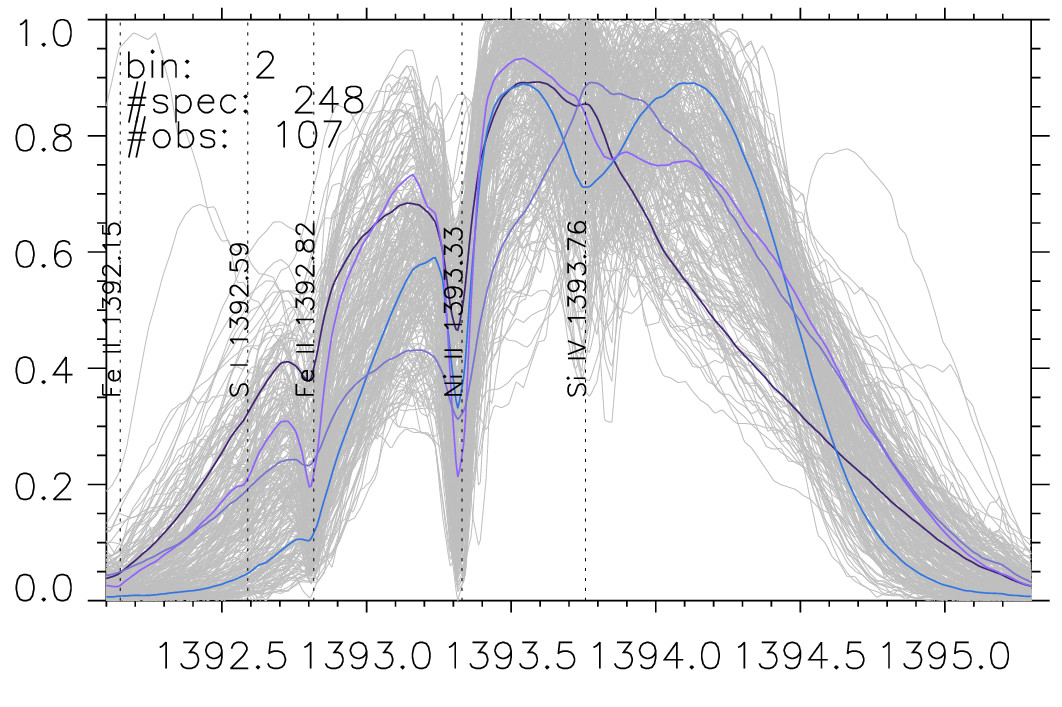}}
    \includegraphics[width=.19\textwidth]{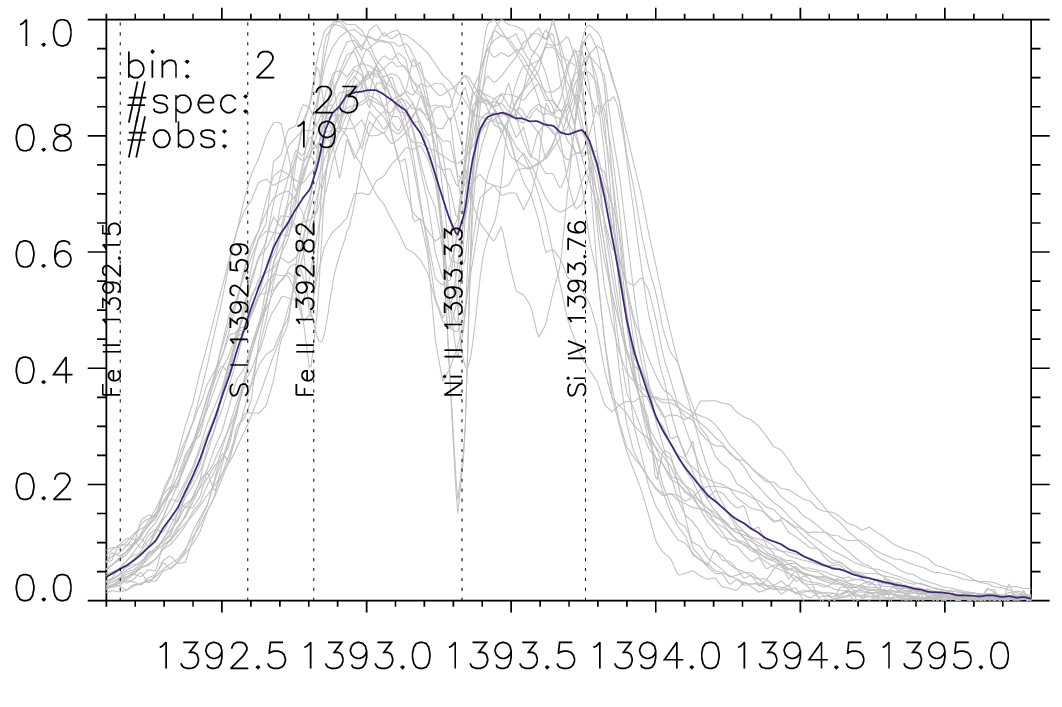}
    \fcolorbox{blue}{white}{\includegraphics[width=.195\textwidth]{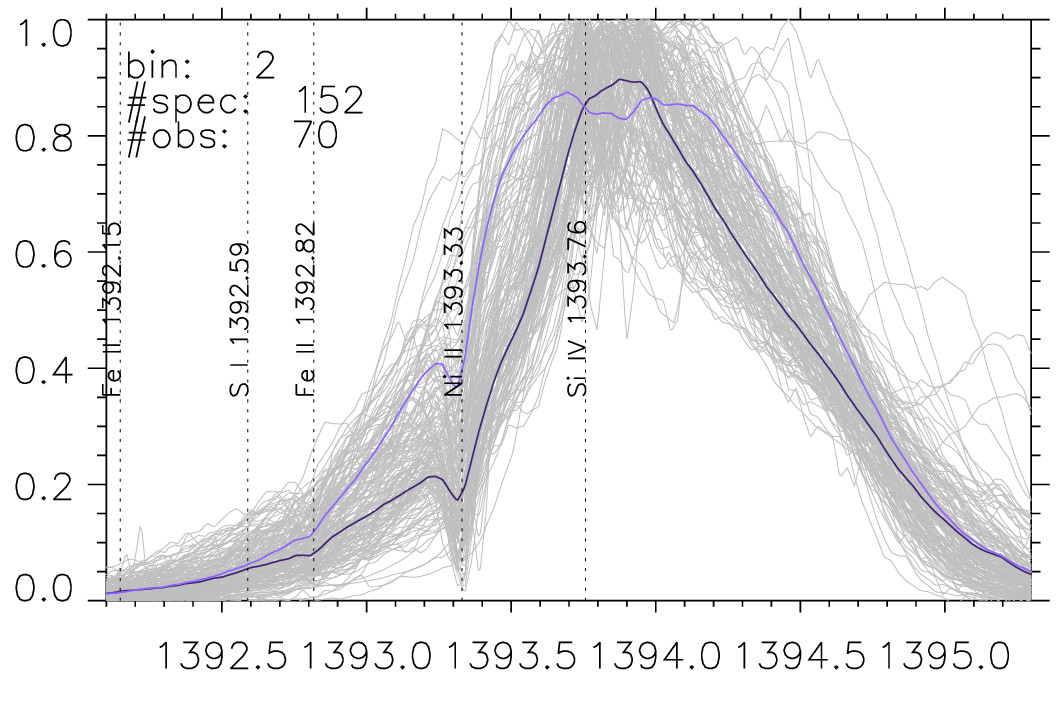}}
   \includegraphics[width=.19\textwidth]{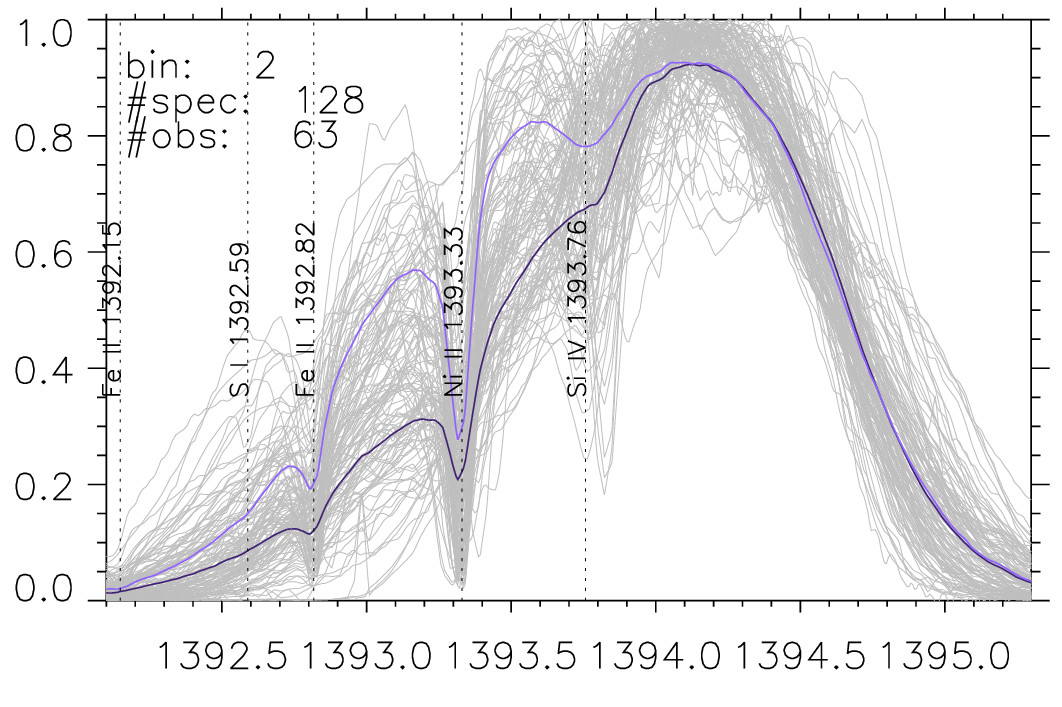}
   \includegraphics[width=.19\textwidth]{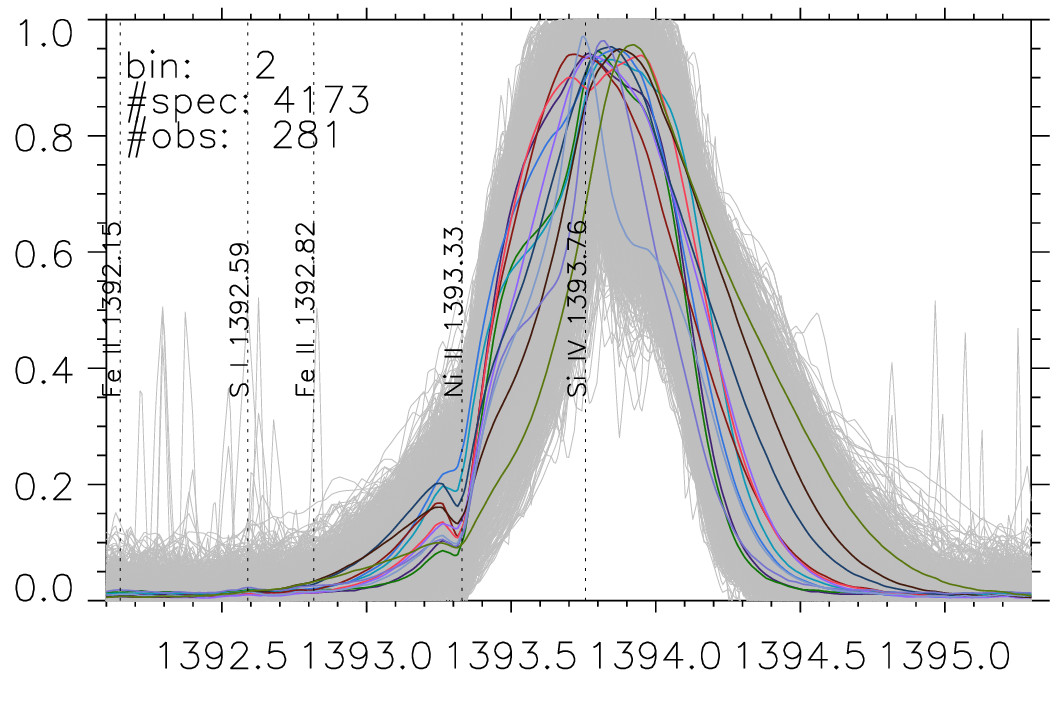}
   \includegraphics[width=.19\textwidth]{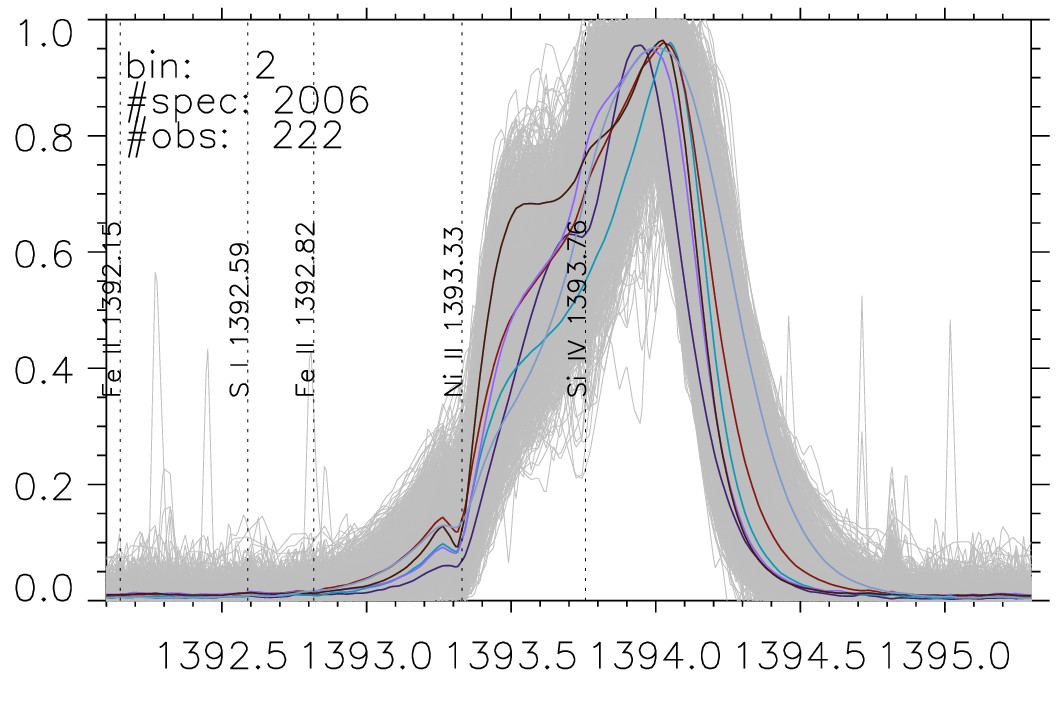}
   \includegraphics[width=.19\textwidth]{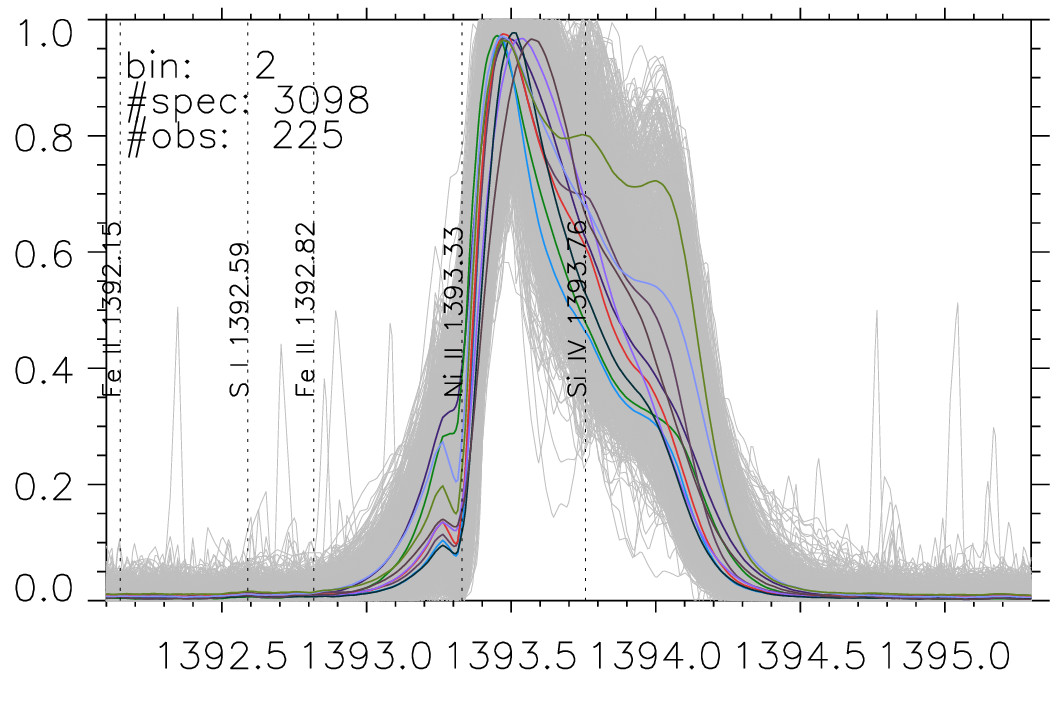}
   \includegraphics[width=.19\textwidth]{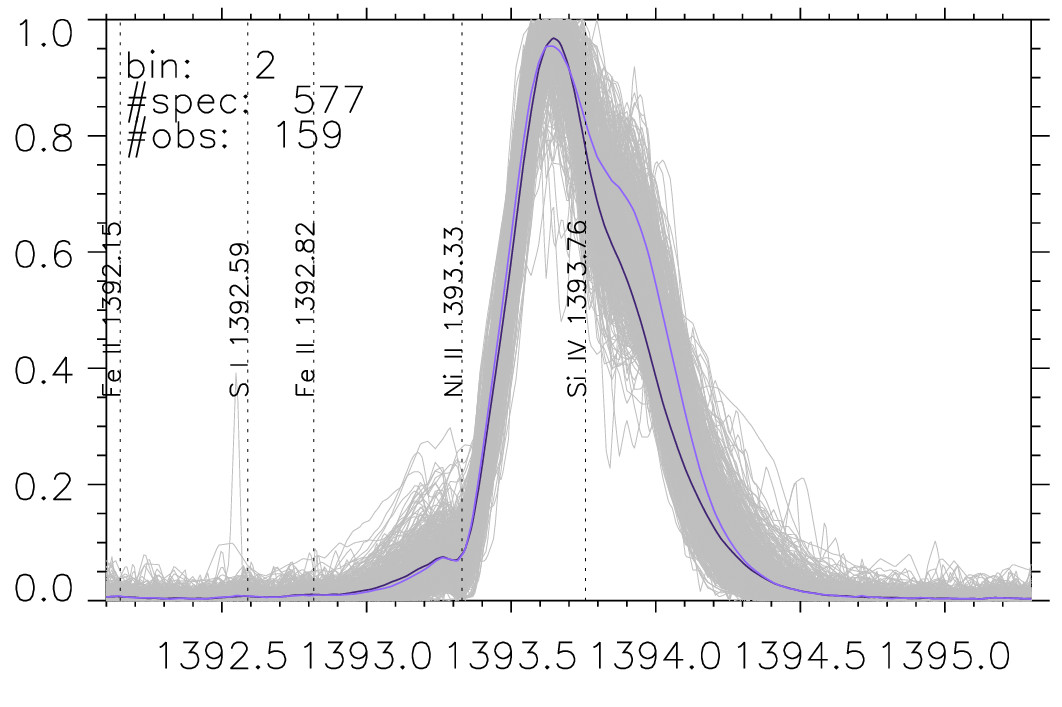}
   \includegraphics[width=.19\textwidth]{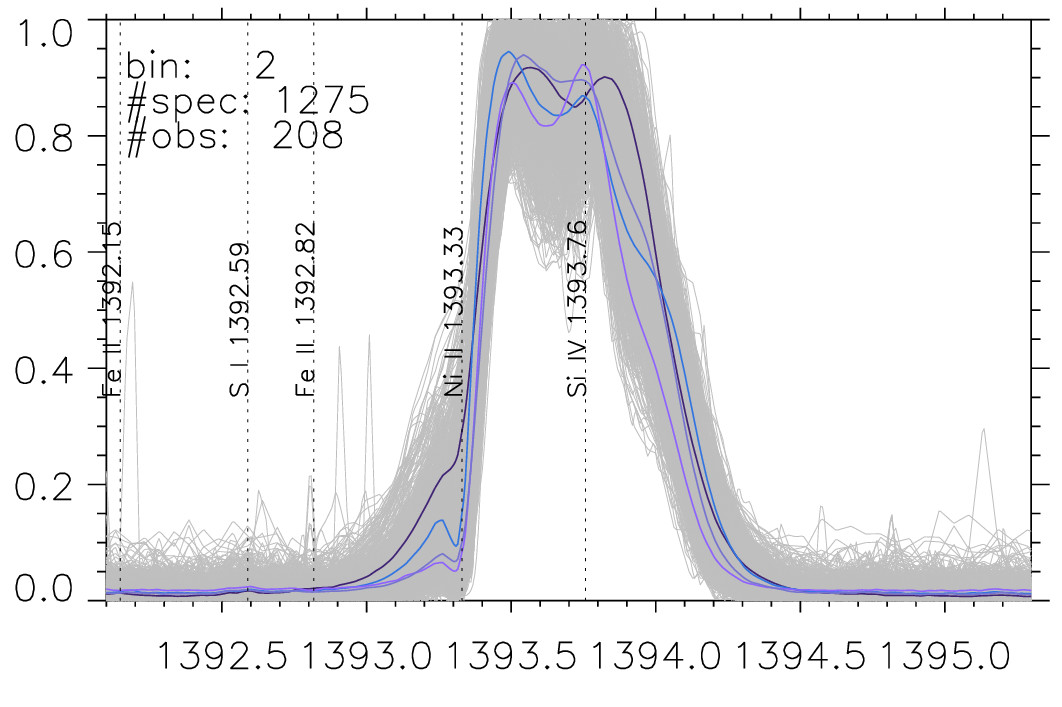}
   \includegraphics[width=.19\textwidth]{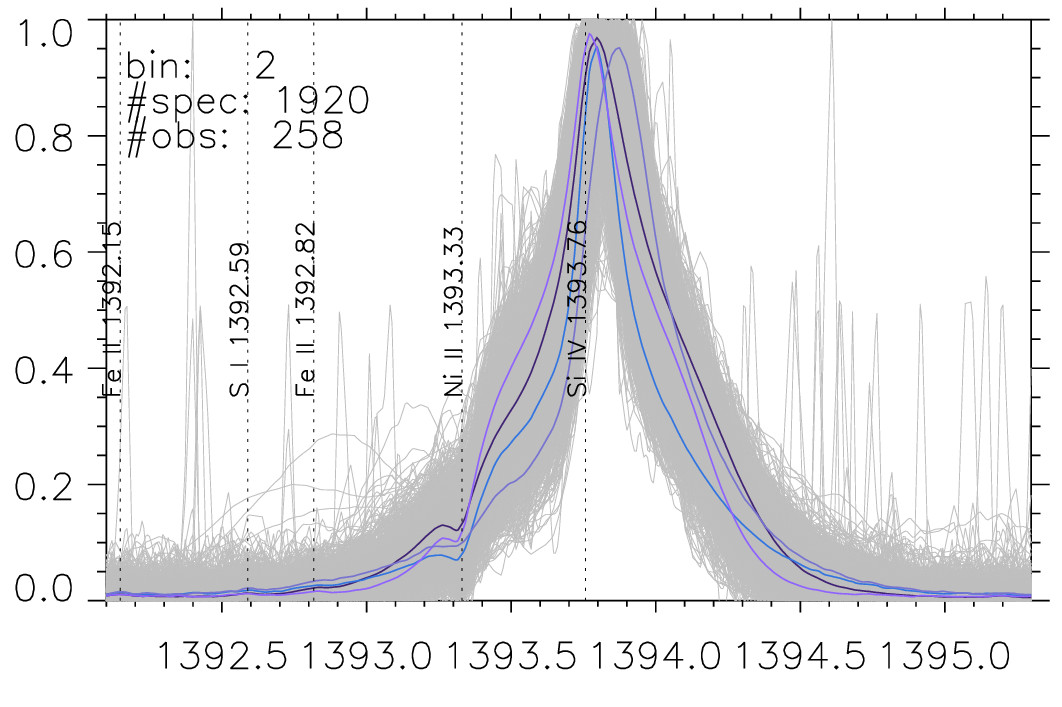}
  \caption{Overview of \ion{Si}{iv} burst spectra (grey) and k-means centroids (colored) that were identified in all observations from 2013-2014. Positions of blend lines are drawn as vertical dotted lines. \#spec indicates how many spectra are drawn in a given panel for a given binning (``bin'') and \#obs in how many of the 287 observations a given type occurred. The yellow-framed burst spectra are analyzed in more detail in Sect.~\ref{ibexamples} and those framed in blue in Sect.~\ref{aia}.}
        \label{fig2}
  \end{figure*}
  
\subsection{AIA and crossalignment}
Because of the varying temporal cadence of the IRIS spectra and slitjaw images (SJI), they are not ideally suited to study the temporal behavior of bursts. We therefore use SDO/AIA \citep{lemenetal2012} data from different passbands, which have constant cadences. Because the bursts were identified in IRIS, a crossalignment between both instruments is necessary, which fortunately was already provided by co-aligned SDO/AIA data cubes as described in ITN32\footnote{\url{https://iris.lmsal.com/itn32/}}. We verified the alignment by crosscorrelating every 5th SJI image (1400 \AA, or 1330 \AA) with its temporally closest AIA 1600 \AA\ image and found that in nearly all cases the offsets were below a couple of arcsec. Because we later use $3\arcsec \times 3 \arcsec$ boxes to create AIA lightcurves, we did not apply any other offsets to the co-aligned data cubes.

We derived the slit position in solar coordinates from the SJI headers (and the pztx keyword from the rasters) and extracted a lightcurve of $\pm$ 30 AIA time steps and $3\arcsec \times 3 \arcsec$ around this position. AIA data have a 12 or 24 s cadence, depending on wavelength, meaning that we obtained lightcurves with a duration of 12 or 24 minutes. If no AIA data were available for a given observation, the lightcurve was omitted, and similarly, if there were variations in the exposure time in that timeframe, which often led to saturated lightcurves, the data point was omitted.

This resulted in a data set consisting of $101337 \times 9$ light curves covering the wavelengths 94, 131, 171, 193, 211, 304, 335, 1600, 1700 \AA. Approximately 1\% of these light curves were excluded due to the above-mentioned criteria.

\subsection{Data processing}

The data processing follows the methods developed by \citet{panosetal2021} and \citet{panoskleint2021}. Here we only briefly summarize the general idea and the steps, while the calculations can be found in the aforementioned papers. Our goal is to classify co-occurring spectra and lightcurves to answer the question: ``if a burst is observed in \ion{Si}{iv}, how does it propagate throughout the solar atmosphere and can it reach coronal temperatures/heights''. Considering the variety of shapes of burst spectra (see Fig.~\ref{fig1}), we combined the 321 IB centroids after a visual inspection into 25 meta groups to better compare similar spectral shapes. 
The result is shown in Fig.~\ref{fig2} where the colored spectra are k-means centroids, and similar-looking ones were grouped into one plot. The grey spectra depict all IB spectra belonging to a given category and a binning of 2 (each spectrum drawn is the average of its two most-similar spectra) was applied for display reasons. The ``\#obs'' label denotes how many observations are represented by a given spectral type and it can be seen that all these spectral shapes occur in many different observations. This means that bursts share similar properties, at least in \ion{Si}{iv}, throughout different observations.

It would now be possible to simply plot all co-occurring spectra in other lines or lightcurves, but the disadvantage is that each of them is slightly different, meaning the mapping  \ion{Si}{iv}$\rightarrow$\ion{Mg}{ii} may have thousands of connections, each with a weight one (single occurrence). But for a statistical analysis it makes sense to determine the frequency of which spectral types occur simultaneously, meaning that there need to be discrete groups for the mapping.
To create such groups, we independently classified all other IRIS spectral lines (normalized) and AIA lightcurves (both not normalized and normalized) using k-means with $k=200$ and a variety of $k$ for the normalized AIA lightcurves to investigate potential biases. This allowed us to derive the connections between groups, as depicted in Fig.\,6 of \citet{panoskleint2021}.

\section{Bursts in different wavelengths} 

\subsection{IRIS wavelengths}\label{ibexamples}

\begin{figure*}[p!]
\begin{center}
\includegraphics[width=0.52\textwidth]{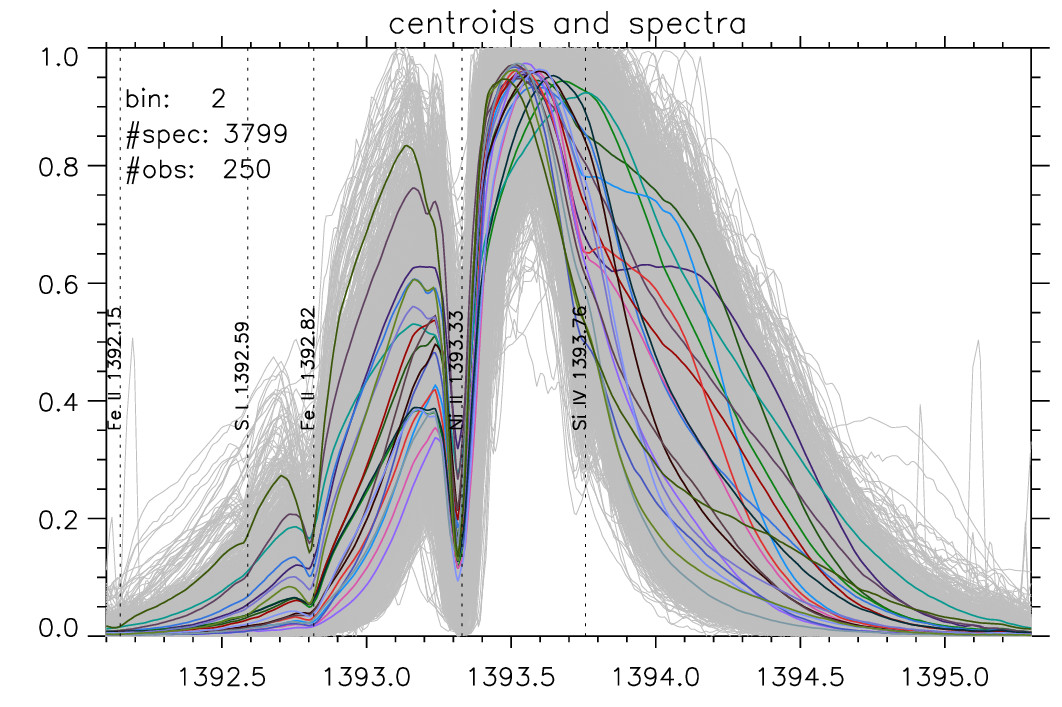}
\includegraphics[width=0.49\textwidth]{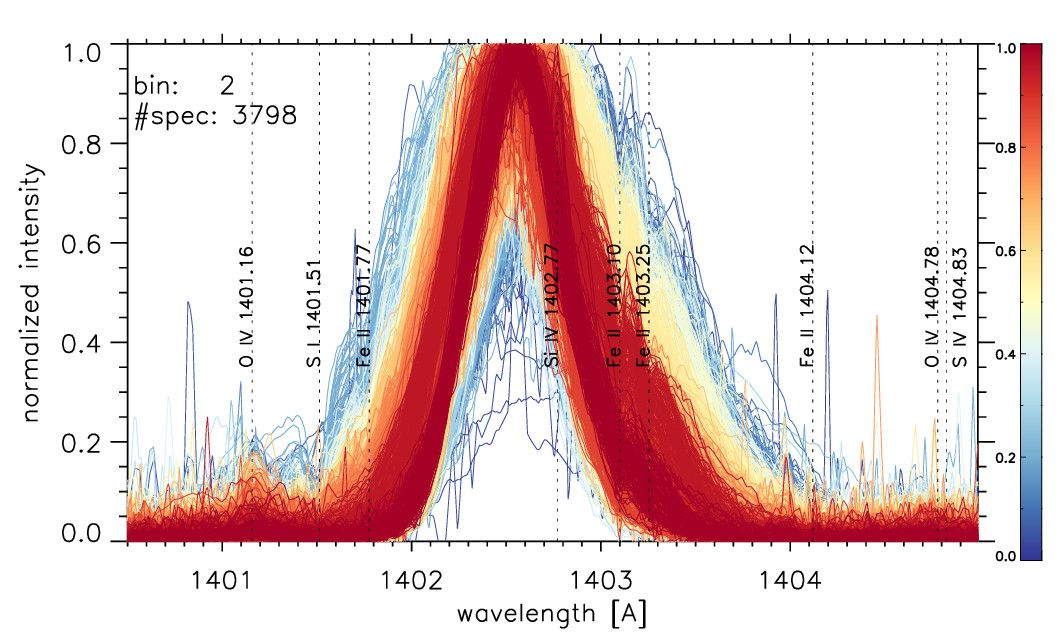}
\includegraphics[width=0.49\textwidth]{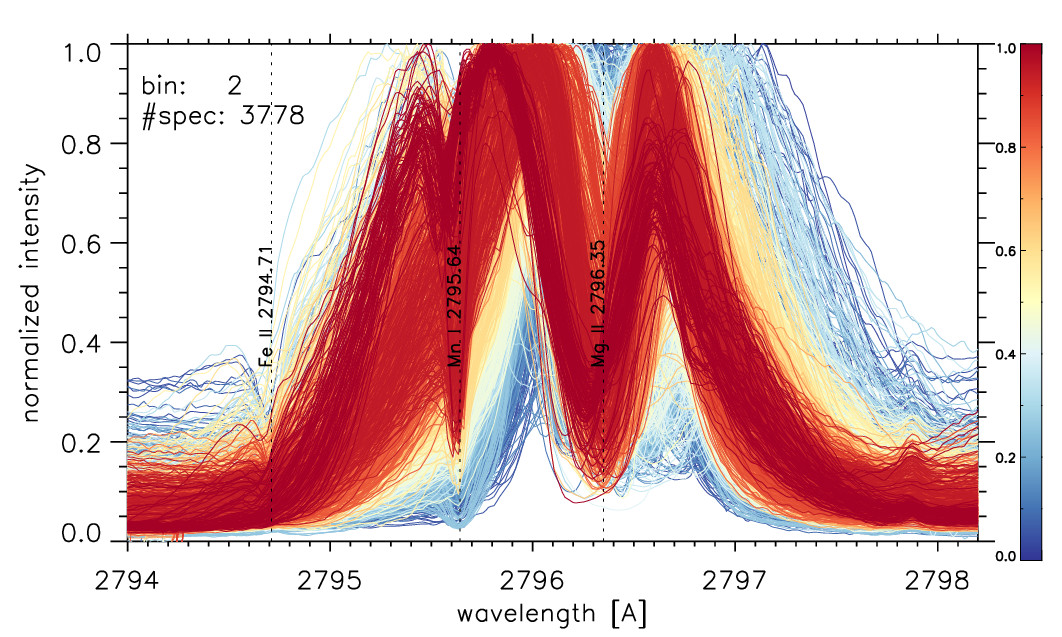}
\includegraphics[width=0.49\textwidth]{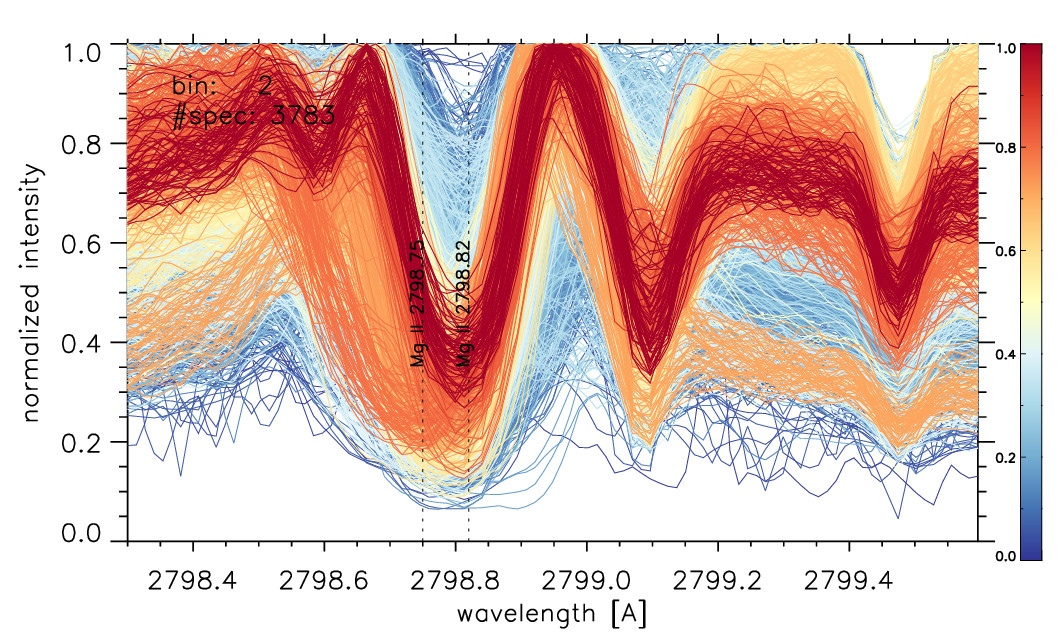}
\includegraphics[width=0.49\textwidth]{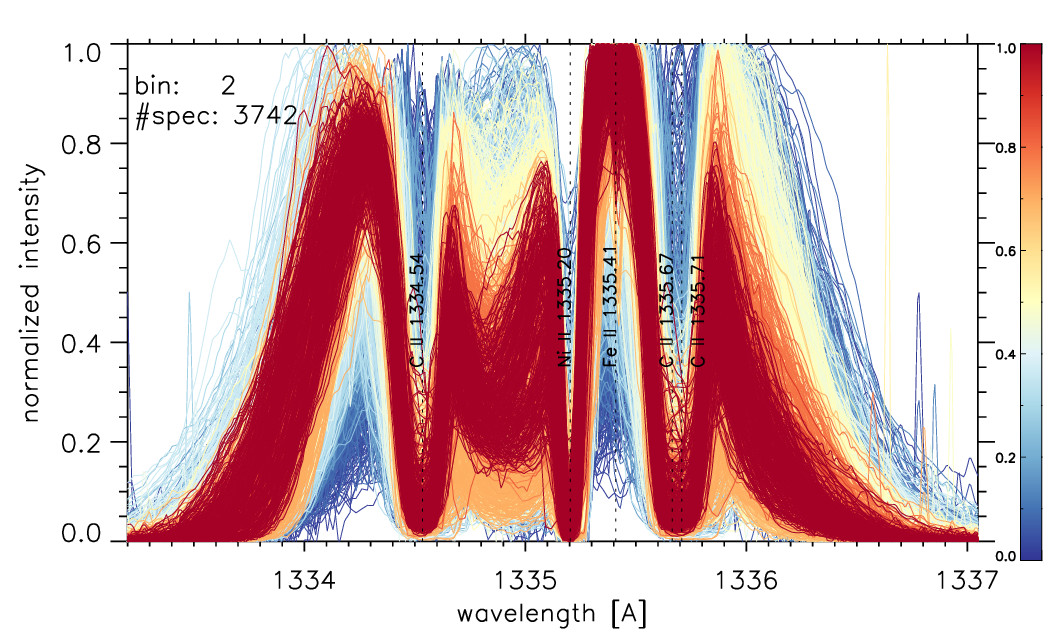}
\includegraphics[width=0.49\textwidth]{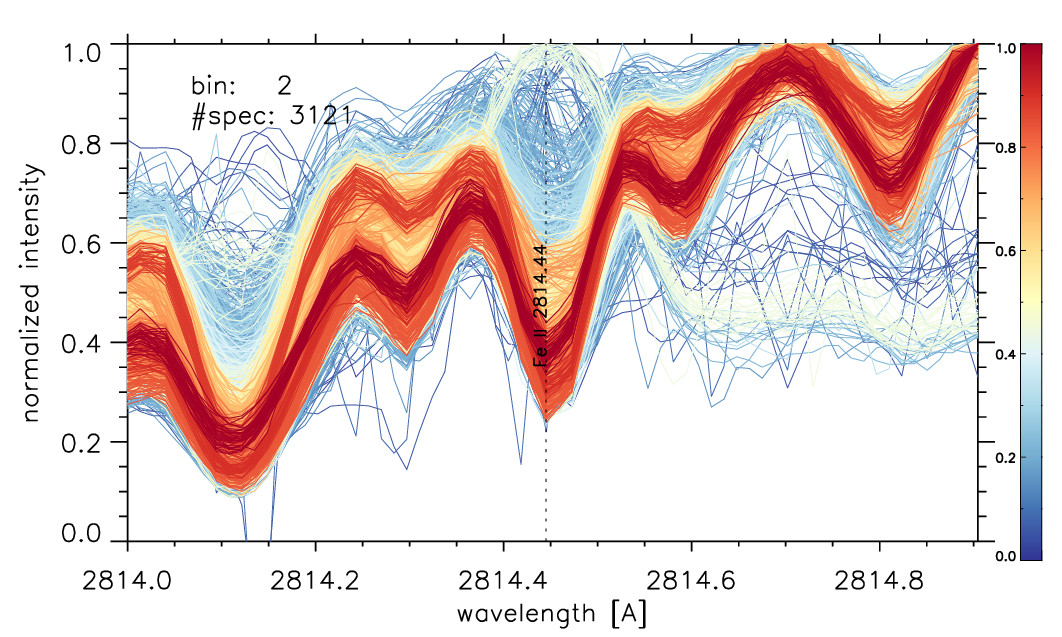}
\includegraphics[width=0.49\textwidth]{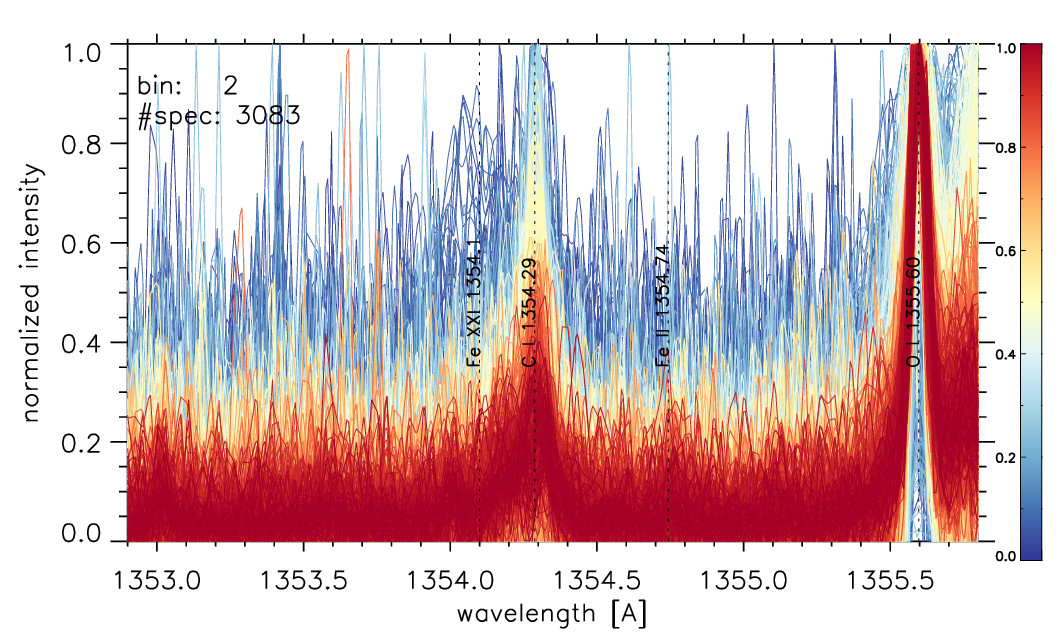}

\caption{Some of the deepest absorption line blends observed in burst spectra. The top panel indicates which \ion{Si}{iv} 1393.76 spectra were selected (grey) and their centroids (colored). The lower panels show the co-occurring spectra, color-coded by probability, red being the most probable.}\label{deepdip1}
\end{center}
\end{figure*}

\begin{figure*}[!p]
\begin{center}
\includegraphics[width=0.52\textwidth]{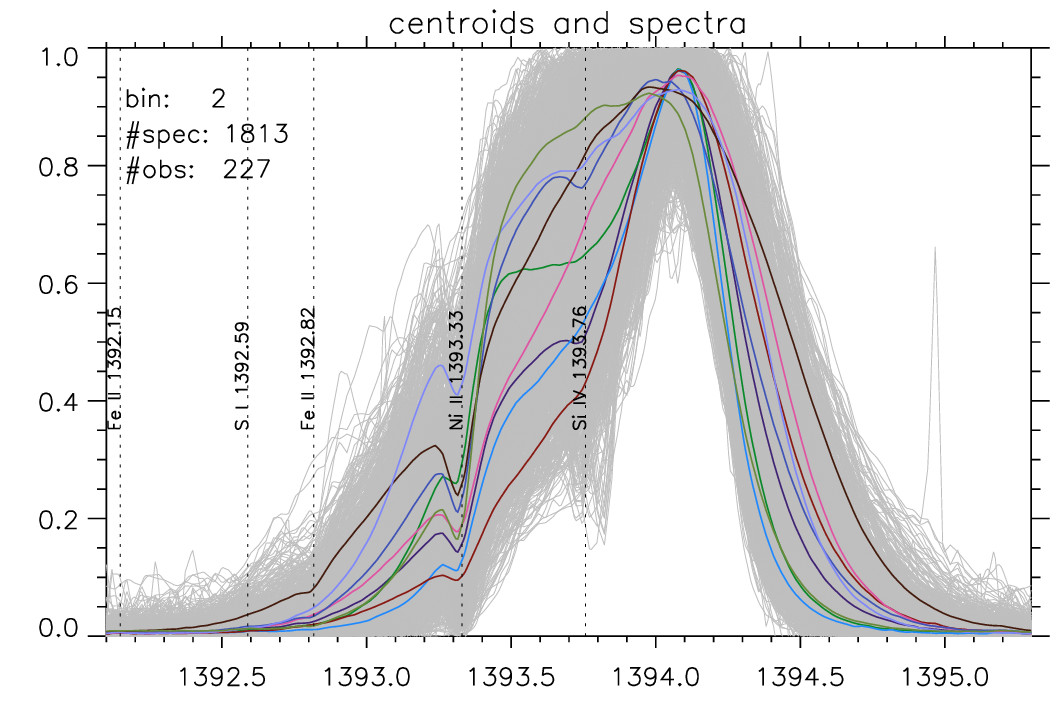}
\includegraphics[width=0.49\textwidth]{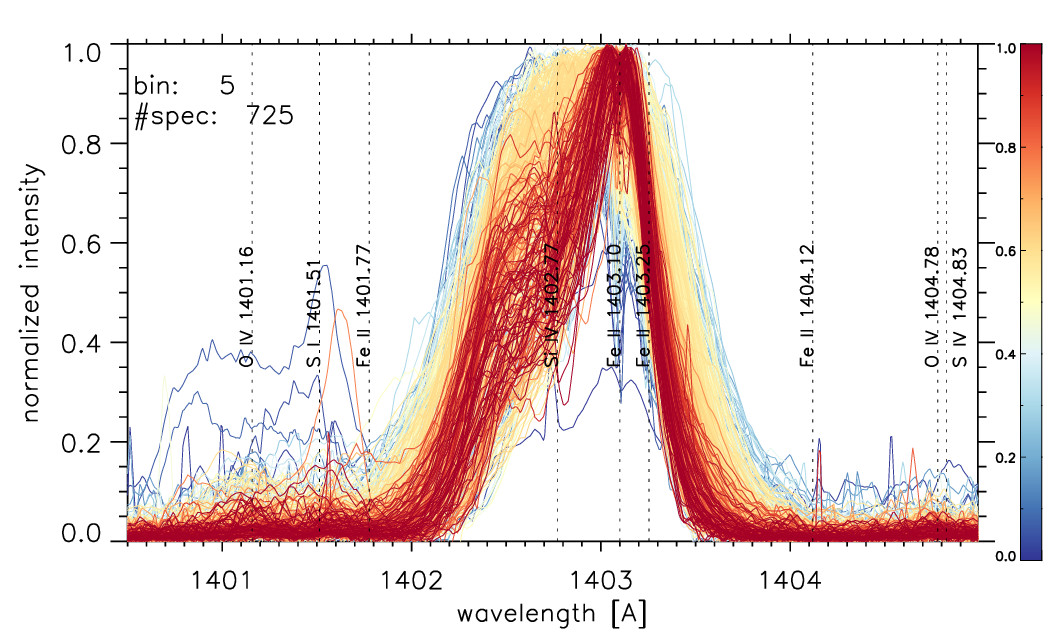}
\includegraphics[width=0.49\textwidth]{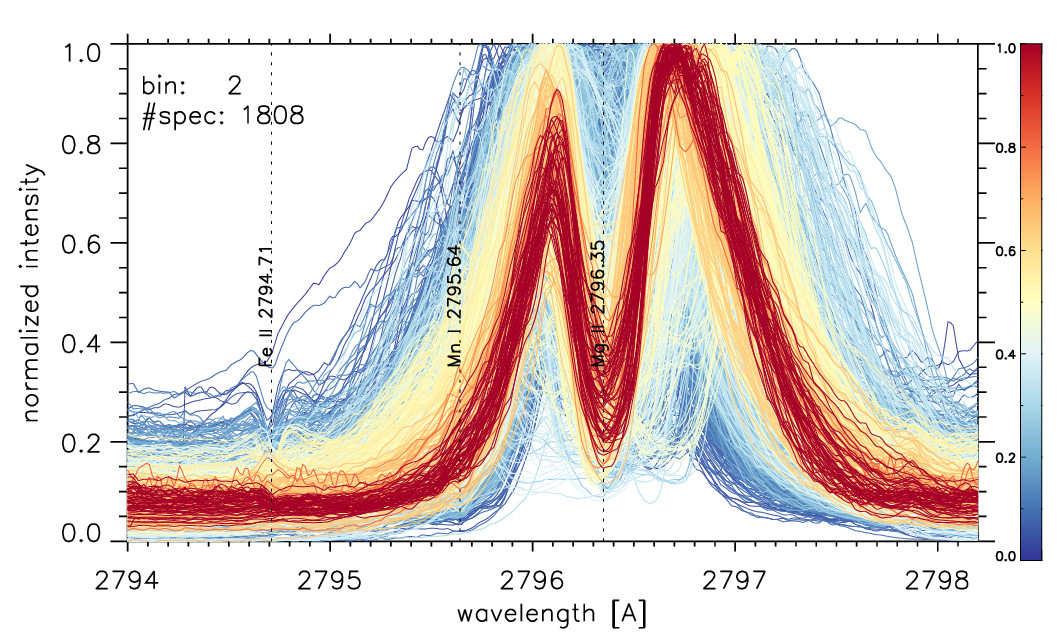}
\includegraphics[width=0.49\textwidth]{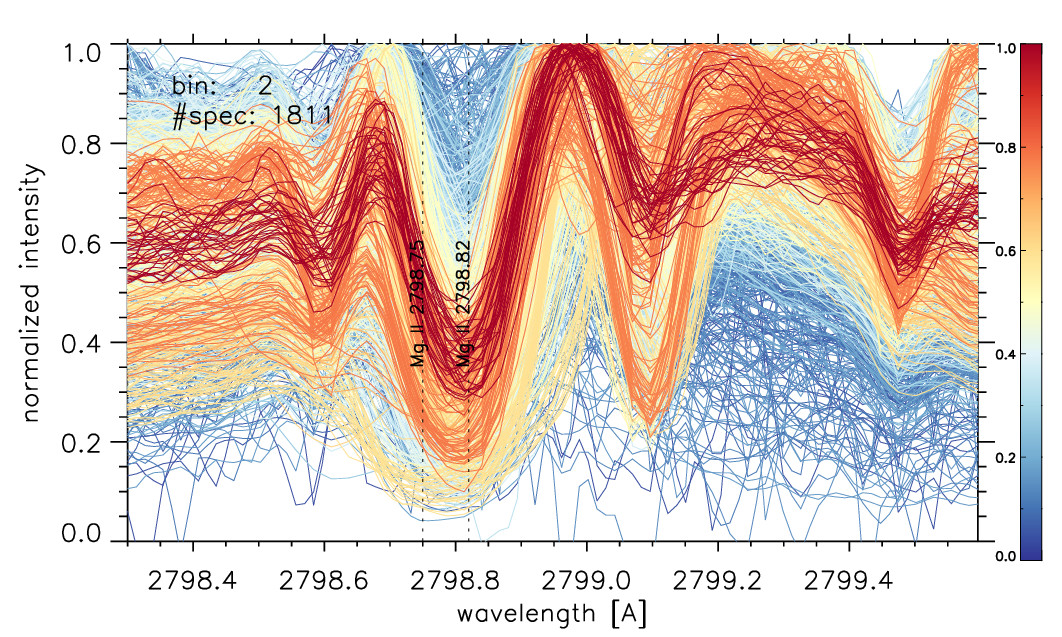}
\includegraphics[width=0.49\textwidth]{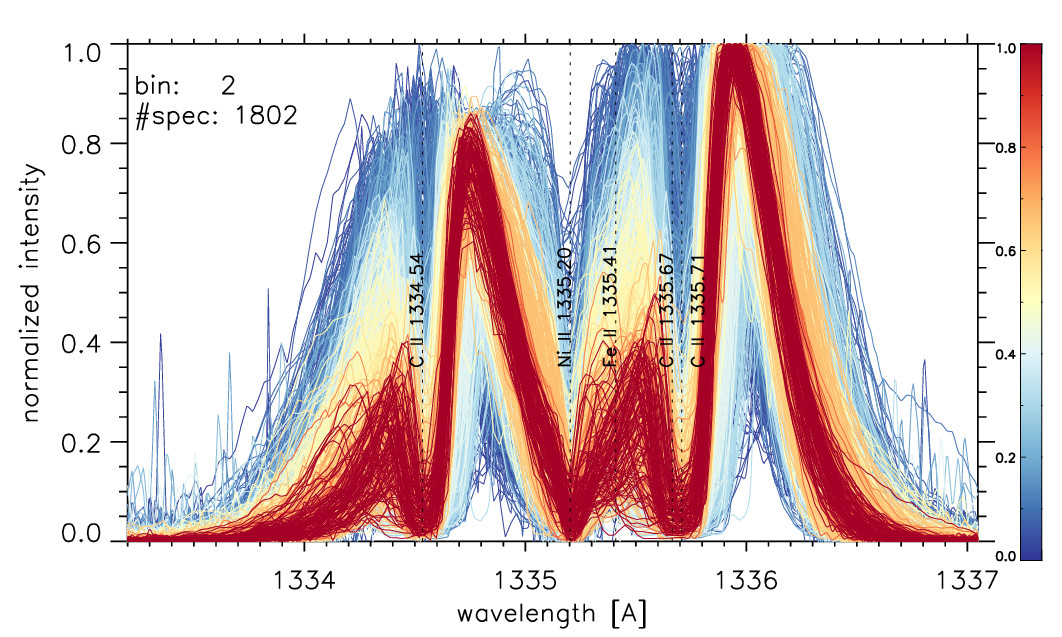}
\includegraphics[width=0.49\textwidth]{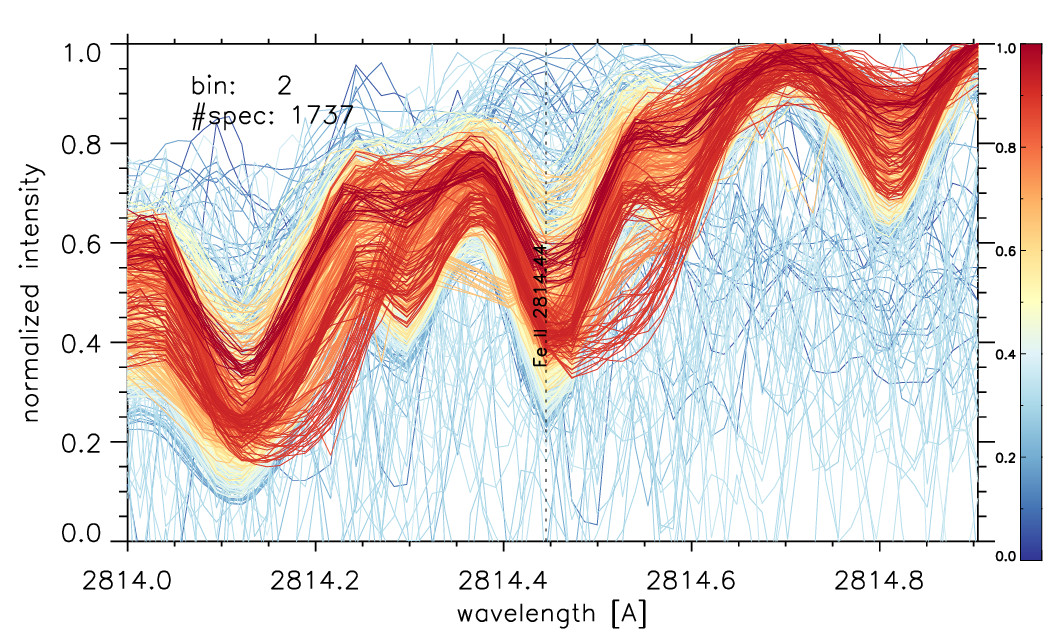}
\includegraphics[width=0.49\textwidth]{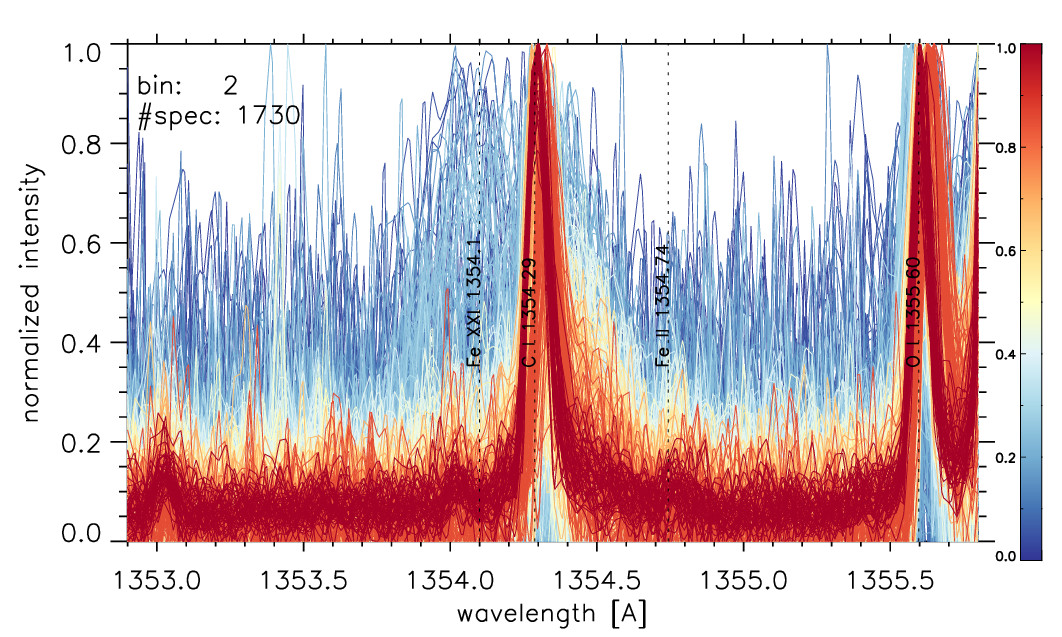}
\caption{Burst spectra with pronounced downflows.}\label{broadbase5}
\end{center}
\end{figure*}

\begin{figure*}[!p]
\begin{center}
\includegraphics[width=0.52\textwidth]{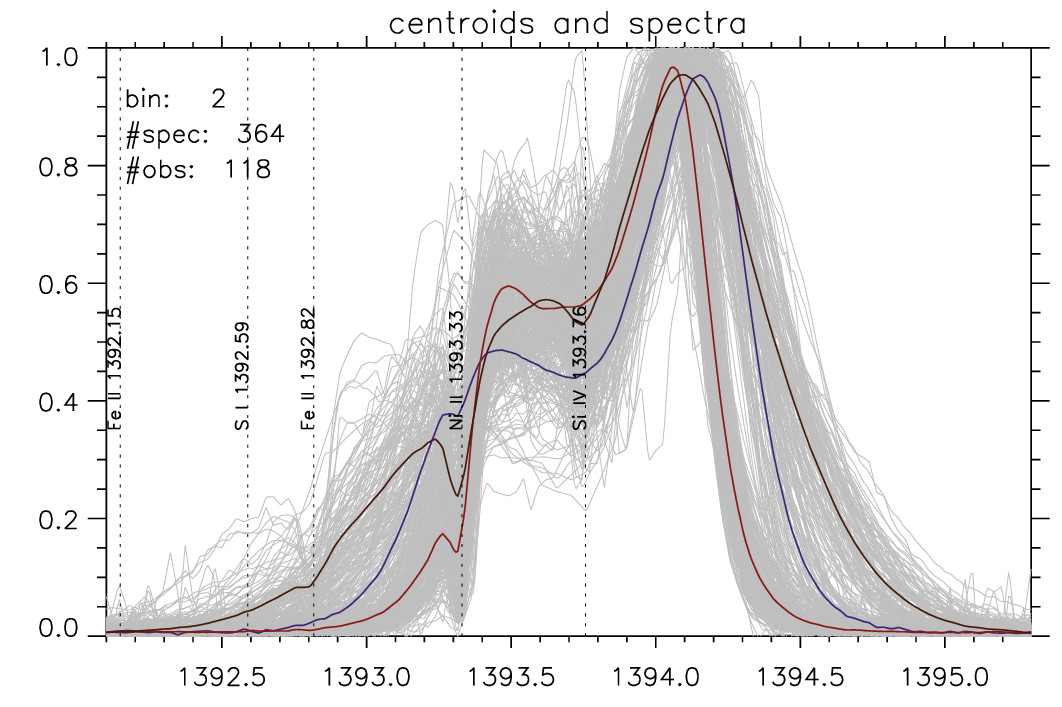}
\includegraphics[width=0.49\textwidth]{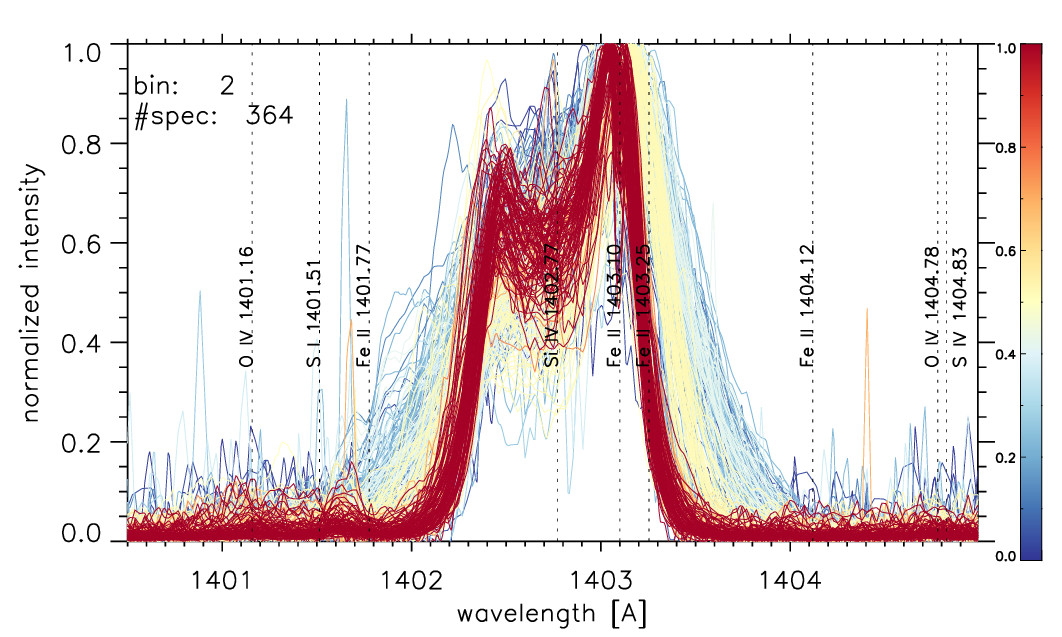}
\includegraphics[width=0.49\textwidth]{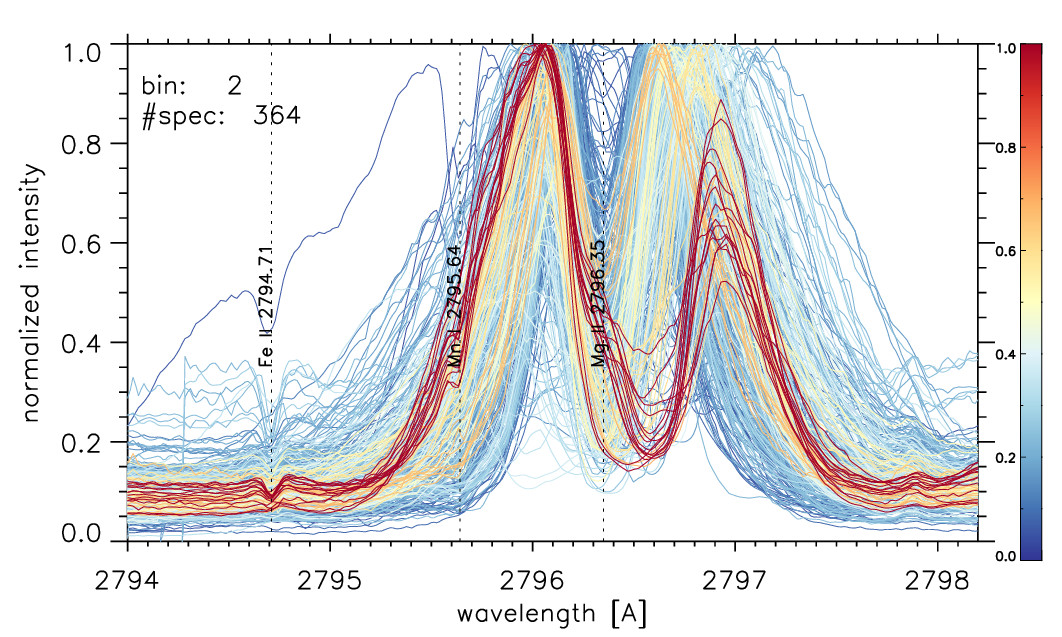}
\includegraphics[width=0.49\textwidth]{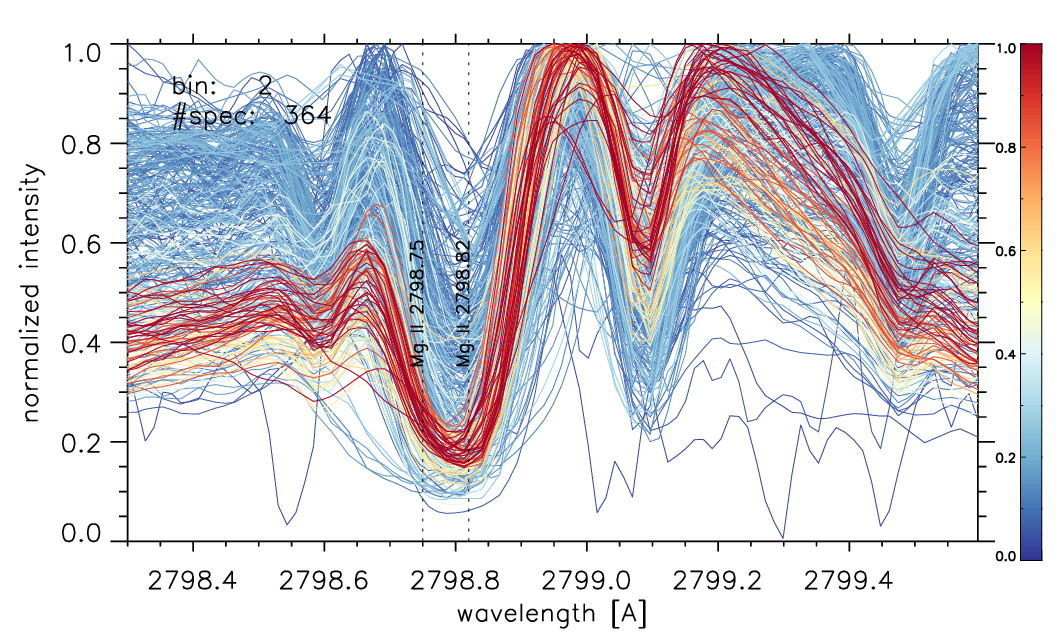}
\includegraphics[width=0.49\textwidth]{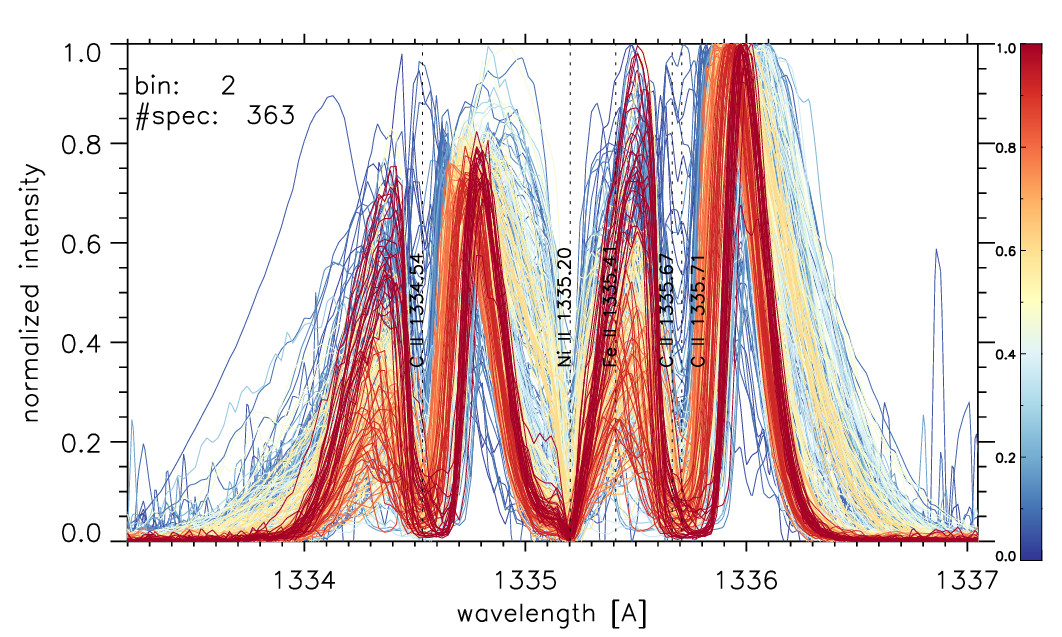}
\includegraphics[width=0.49\textwidth]{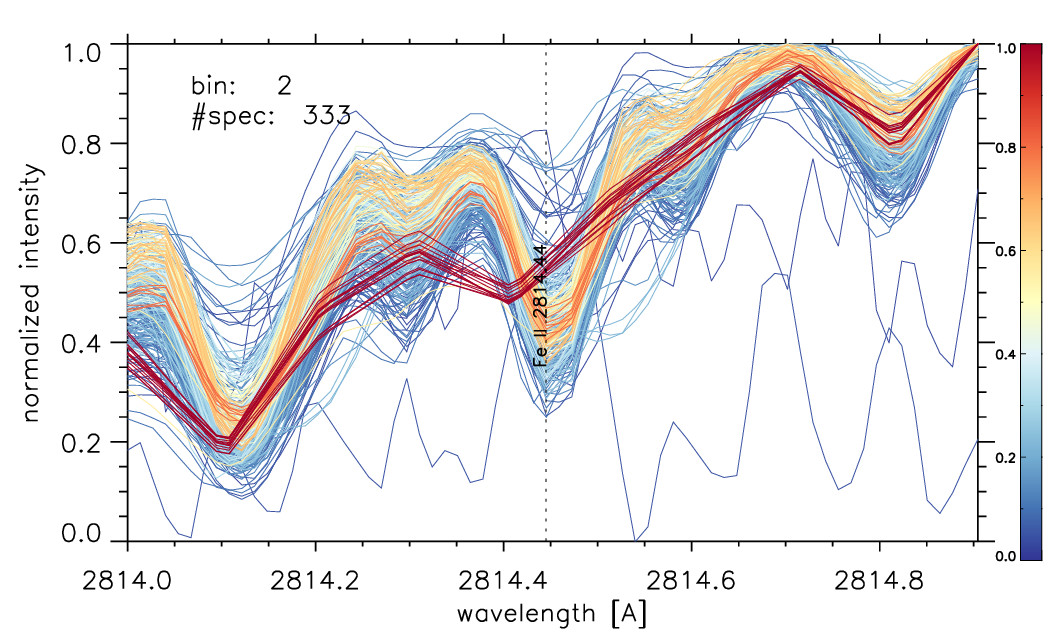}
\includegraphics[width=0.49\textwidth]{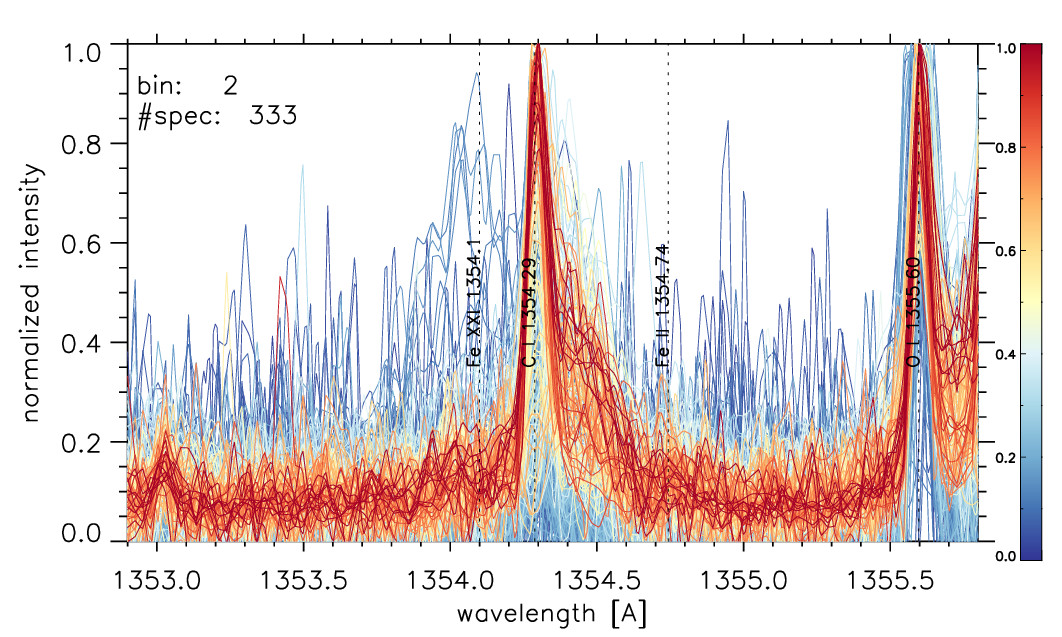}
\caption{Burst spectra with even stronger downflows than Fig~\ref{broadbase5}.}\label{centerdip4}
\end{center}
\end{figure*}

\begin{figure*}[!p]
\begin{center}
\includegraphics[width=0.52\textwidth]{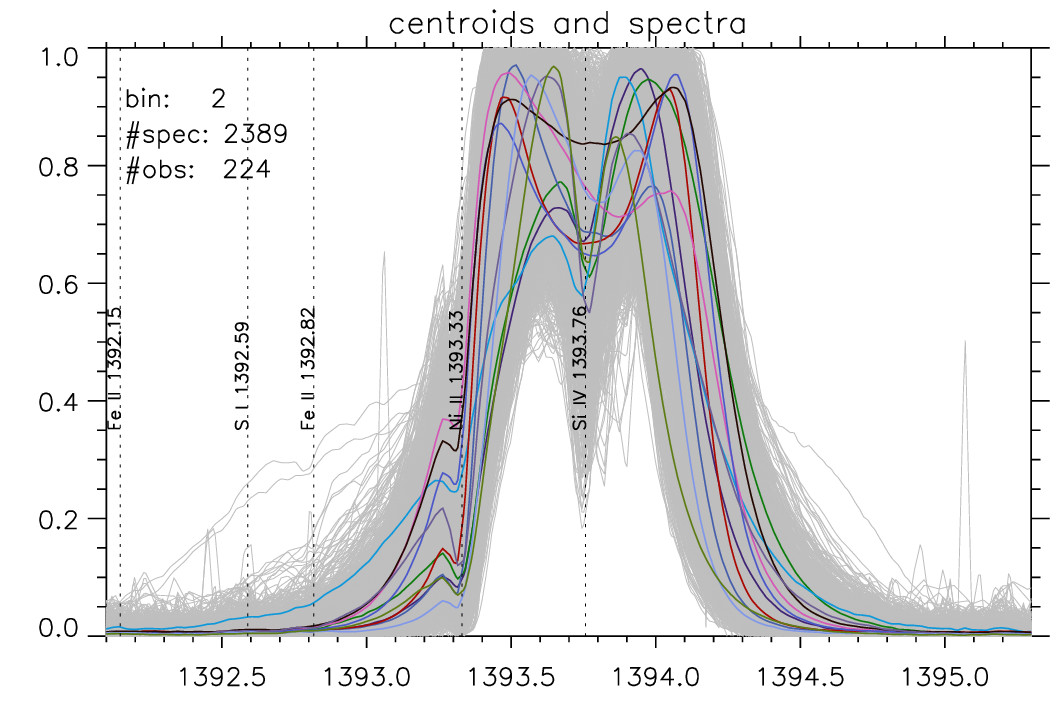}
\includegraphics[width=0.49\textwidth]{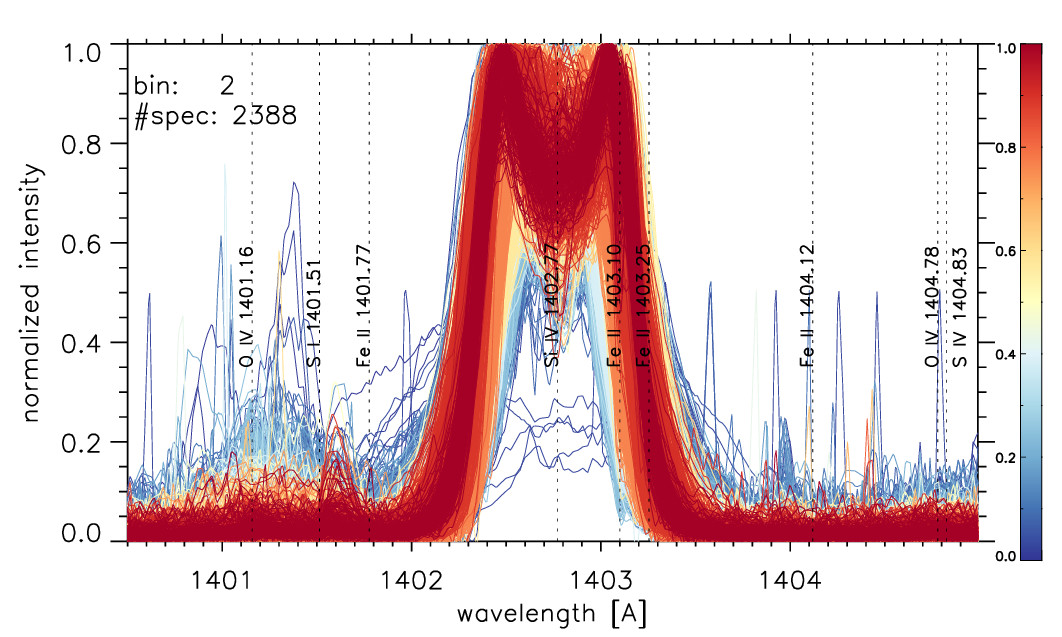}
\includegraphics[width=0.49\textwidth]{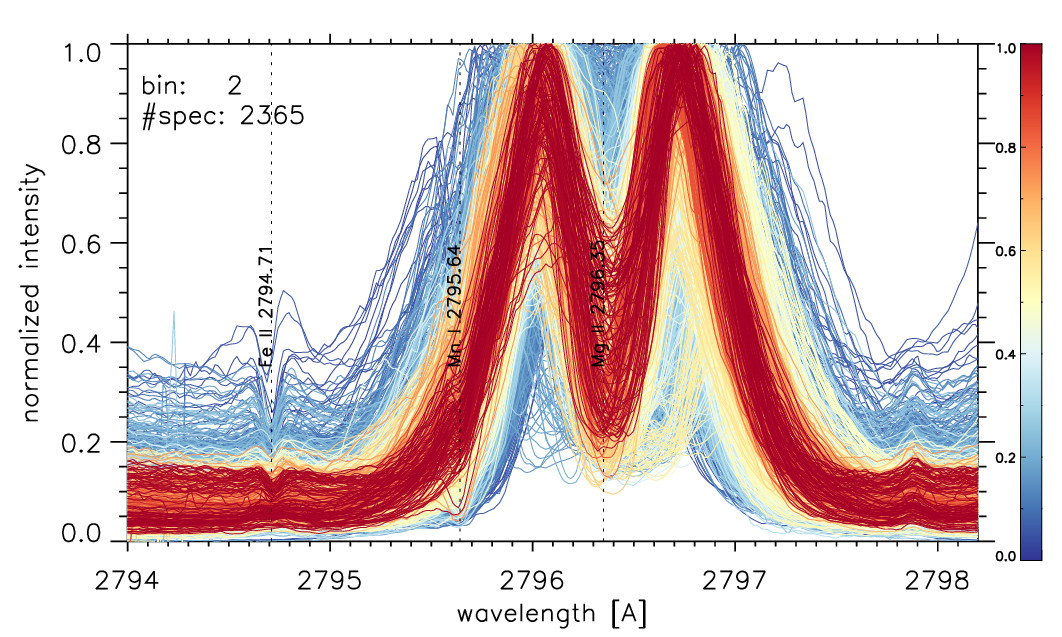}
\includegraphics[width=0.49\textwidth]{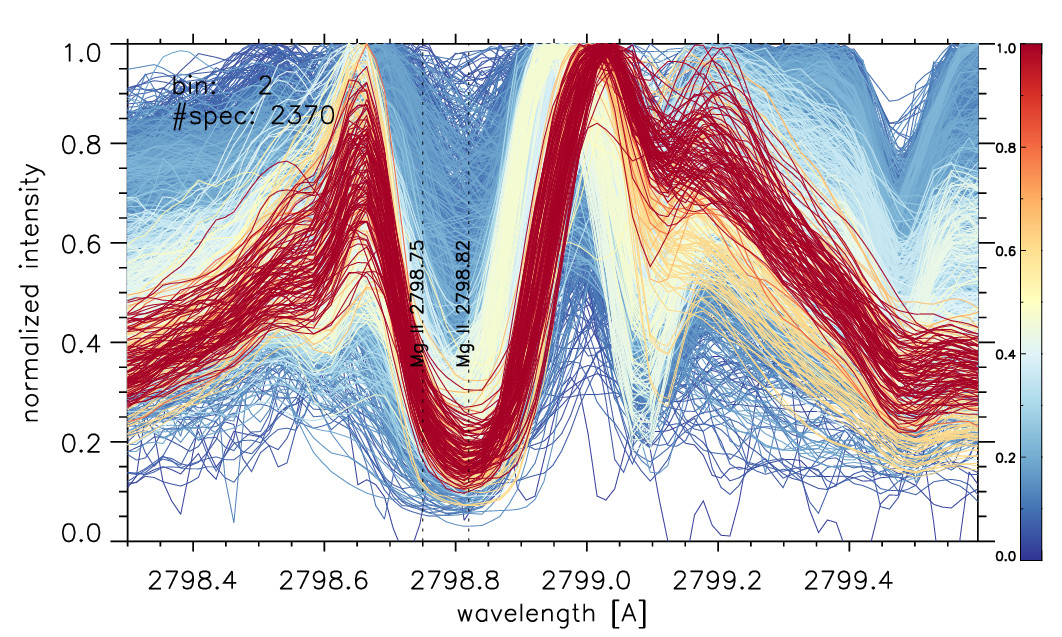}
\includegraphics[width=0.49\textwidth]{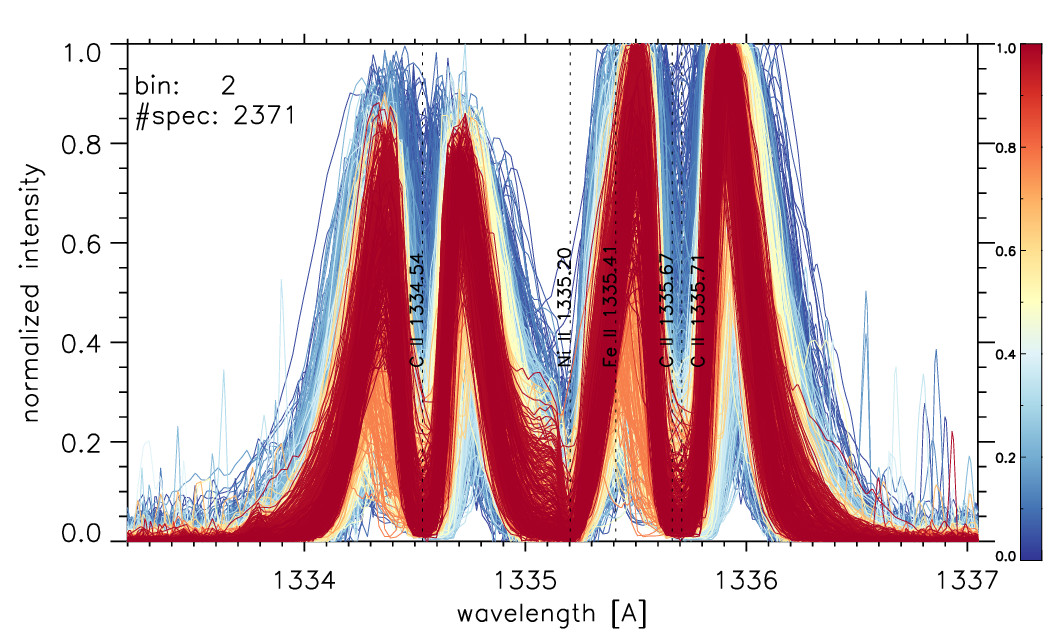}
\includegraphics[width=0.49\textwidth]{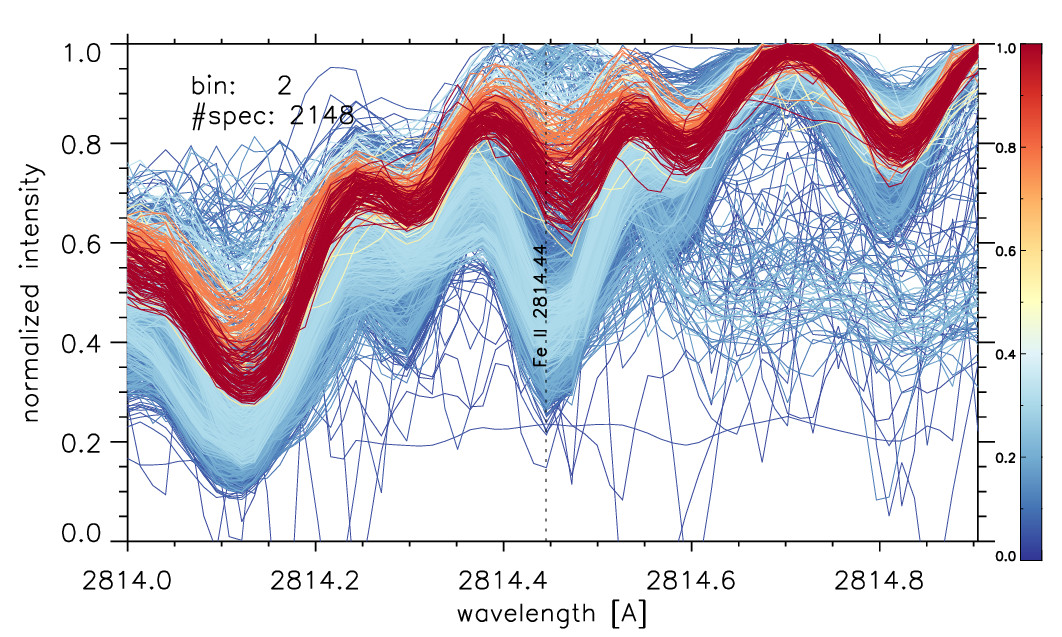}
\includegraphics[width=0.49\textwidth]{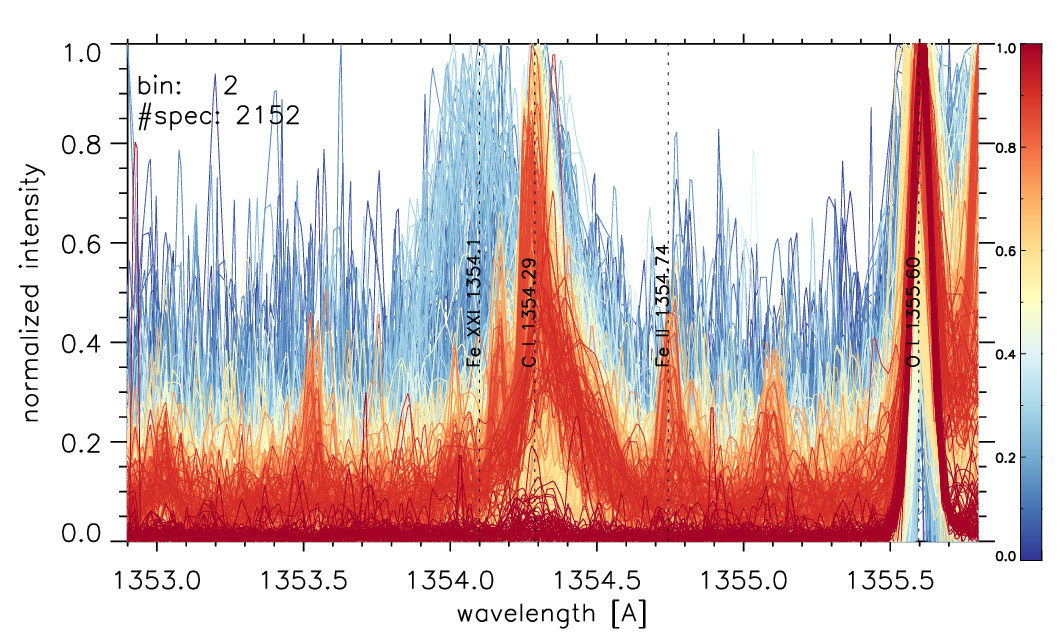}
\caption{Burst spectra that resemble the first example from \citet{peteretal2014science}, which was interpreted as high up- and downward velocities.}\label{centerdip1}
\end{center}
\end{figure*}

We now investigate four interesting types of burst spectra associated with the yellow-framed panels in Fig.~\ref{fig2}, specifically concentrating on their multi-thermal properties. These include spectra with the deepest absorption in the blend lines, cases of highly asymmetric spectra (probably indicating flows), and more symmetric spectra with a central dip that in the past were interpreted as bi-directional flows. The results are depicted in Figure~\ref{deepdip1} -- \ref{centerdip1}, where the top panel shows a magnification of the relevant panel of Fig.~\ref{fig2} and the plots below contain spectra from different lines that are co-occurring with the \ion{Si}{iv} burst spectra. The spectra in these panels are color coded such that the most common/probable spectra are red while the less likely spectra are blue. The probabilities are normalized to the most-frequently occurring spectral type, meaning e.g.\,yellow spectral shapes, which correspond to a probability of 0.5, occur half as often as red ones.

 \begin{figure*} 
  \centering 
   \includegraphics[width=.49\textwidth]{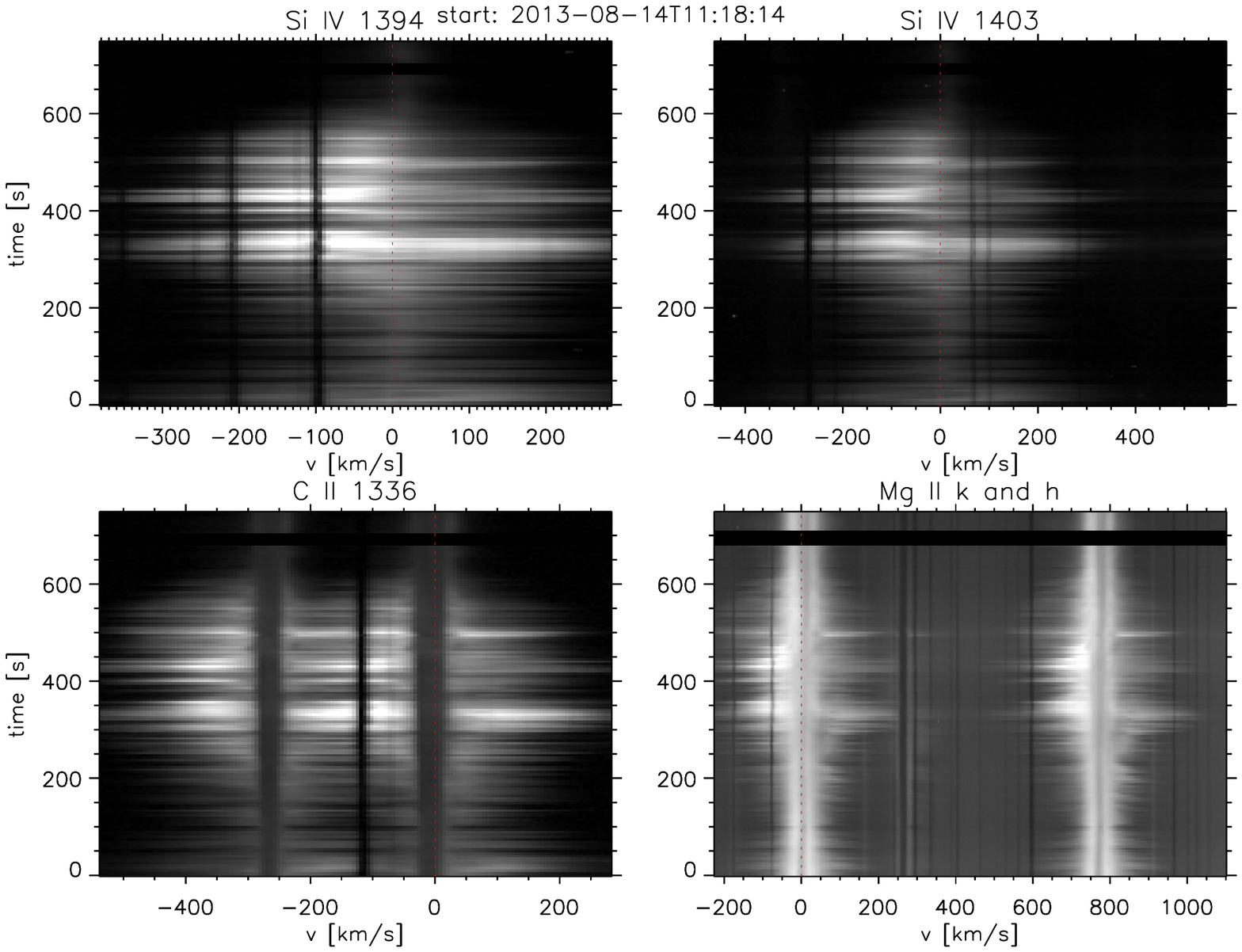}
   \includegraphics[width=.49\textwidth]{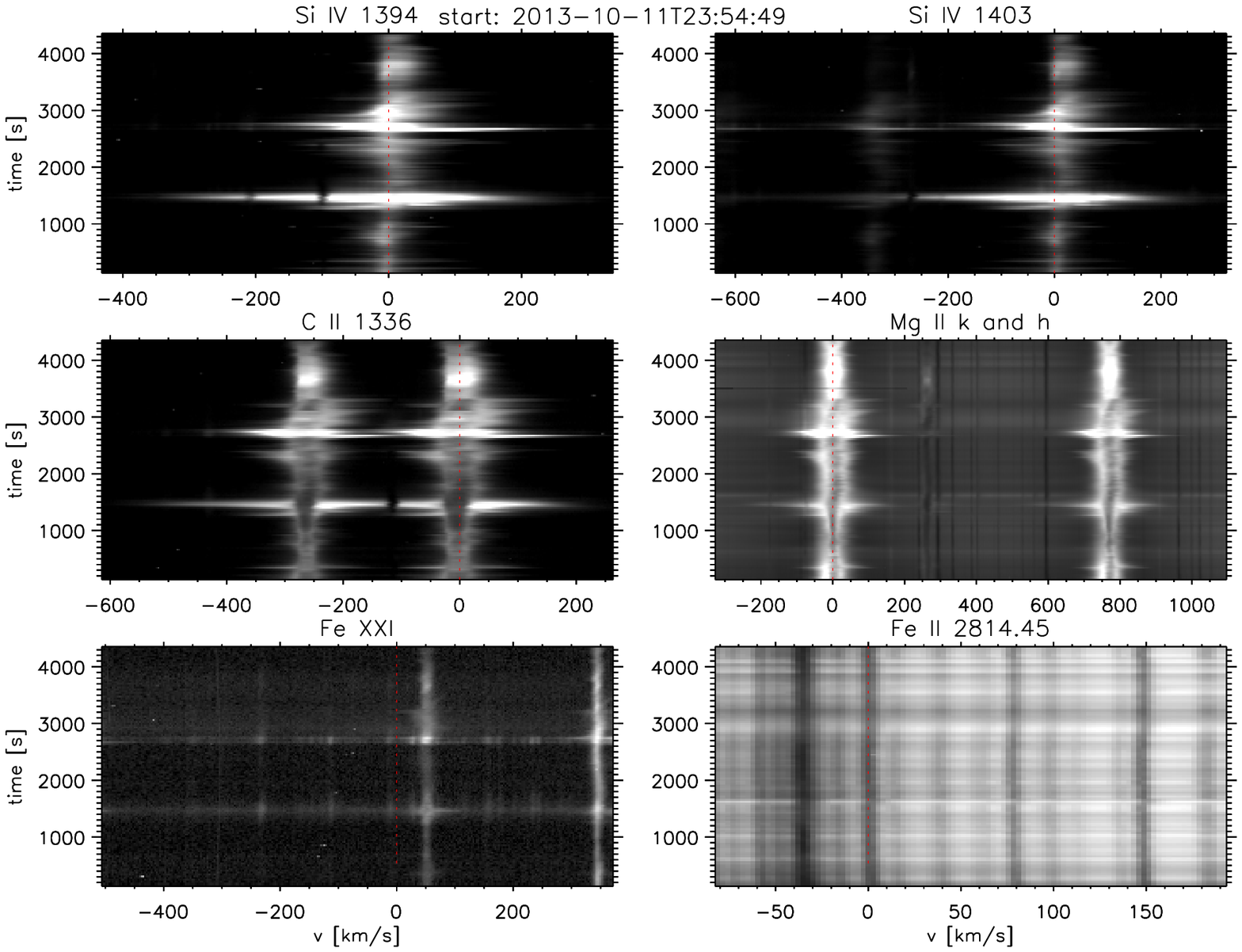}
   \caption{Examples of temporal evolutions of bursts. The horizontal axis is labeled in Doppler velocities, while the vertical axis is given in seconds. Left: Recurring bursts, visible in all spectral lines with absorption line blends. Right: The first burst is an IB, while the second burst ($\sim 20$ min later) seems to occur higher in the atmosphere and thus has no  absorption blend lines.}
        \label{fig6}
  \end{figure*}

Figure~\ref{deepdip1} shows the deep blend absorptions at \ion{Si}{iv} 1393.8 (via \ion{Ni}{ii} 1393.3, \ion{Fe}{ii} 1392.82, hints of \ion{Si}{i} 1392.59), at \ion{Si}{iv} 1402.77 (weakly via \ion{Si}{i} 1401.51, \ion{Fe}{ii} 1401.77, \ion{Fe}{ii} 1403.10, \ion{Fe}{ii} 1403.25, best visible with higher binnings), at \ion{Mg}{ii} 2796.35 (via \ion{Fe}{ii} 2794.71 and \ion{Mn}{i} 2795.64), in the \ion{C}{ii} region all lines show significant absorption (\ion{C}{ii} 1334.54, \ion{Ni}{ii} 1335.20, \ion{Fe}{ii} 1335.41, \ion{C}{ii} 1335.67 and 1335.71), the Mg triplet appear as absorption lines with slightly intensity-enhanced line wings, while in the FUV, the \ion{C}{i} 1354.29 and \ion{O}{i} 1355.60 lines appear in significant emission, which usually does not occur except for flares and energetic events. Considering the depth of the blends, this may suggest that such bursts occur deeper in the atmosphere than others, where significant amounts of cooler material lie above. Most blends are blue-shifted, similarly as observed by \citet{peteretal2014science} and consistent with the interpretation of hot expanding material deep in the atmosphere, pushing the above cooler material upwards. 
This case does not seem to be very rare, considering that it appeared in 250 of the 287 IRIS observations.

Figure~\ref{broadbase5} shows particular spectra that have a strong peak on the red side, probably indicating downflows. This is mirrored in all spectral lines, including the FUV emission lines (\ion{C}{i} 1354.29 and \ion{O}{i} 1355.60). 
The mechanism responsible for such spectra is not entirely clear. 
The peak intensity is offset from line center by about 65 km s$^{-1}$, indicating that hot plasma must be flowing away from the observer. In addition, cool material on top of the burst is moving towards the observer.
There is a hint of \ion{Fe}{xxi} emission in rare cases prompting us to perform the analysis in reverse: we therefore fixed all \ion{Fe}{xxi} emission spectra and analyzed the co-occurring spectra from the other lines.
Only about 1000 of 95000 spectra showed clear \ion{Fe}{xxi} emission and most of them were correlated to Gaussian-type \ion{Si}{iv} spectra that looked narrow in comparison to most burst spectra. Considering their extreme rarity, we omitted a more in depth analysis for this study, which should at some point be undertaken. In summary, \ion{Fe}{xxi}  emission in burst spectra is extremely rare and it is unclear if they are true bursts.

Figure~\ref{centerdip4} shows highly asymmetric spectra, probably due to strong downflows. Particularly noteworthy are the two \ion{Mg}{ii} triplet lines near  2798.8 \AA, where downflowing plasma has created a broad red wing, which is visible in most other spectral lines as well.

Lastly, Figure~\ref{centerdip1} contains spectra that appear most similar to the first reported IB in \citet{peteretal2014science}, which was nearly symmetric about line center and had broad wings on either side as well as a dip at line center. 
The spectrum was interpreted as high up- and downflows, arising from the plasma that is expanding up- and downwards in the atmosphere after a magnetic reconnection. We can now confirm the spectral properties statistically, with similar line shapes occurring in \ion{Mg}{ii} , \ion{C}{ii}, and \ion{Si}{iv}. Our spectral group includes a few asymmetric spectra (stronger downflows), which may explain the appearance of downflows observed in both the \ion{Fe}{ii} 2814.44 and \ion{C}{i} 1354.29 lines. 
This group may also contain traces of \ion{Si}{iv} spectra whose central reversals are due to self-absorption rather than flows \citep[see][]{yanetal2015}. Self-absorption in general is characterized by a relatively narrow dip near line center and the spectra cannot be reasonably fitted by two Gaussians (i.e. bi-directional flows).
The majority of self-absorbed \ion{Si}{iv} spectra, including the ones reported by the above cited publication, were more likely classified into the three groups on the right of the first row of Fig.~\ref{fig2}.

In all our examples, the major IRIS spectral lines (\ion{Mg}{ii}, \ion{C}{ii}, \ion{Si}{iv}) showcase similar properties (e.g.\,flows and line shapes) and reflect a similar behavior of blend lines, indicating that bursts are multi-thermal in nature, and minimally span a temperature range of $10^4 - 10^5$ K.
They most likely do not reach $10^7$ K (\ion{Fe}{xxi}), at least not simultaneously, or in a sufficient volume to cause emission above the noise. All these spectra differ significantly from quiet Sun, sunspot, or flare spectra \citep[see e.g.,][]{iris2014, tianetal2014, kleintetal2014, rubiodacostakleint2017, panoskleint2021}.

 \begin{figure} 
  \centering 
   \includegraphics[width=.49\textwidth]{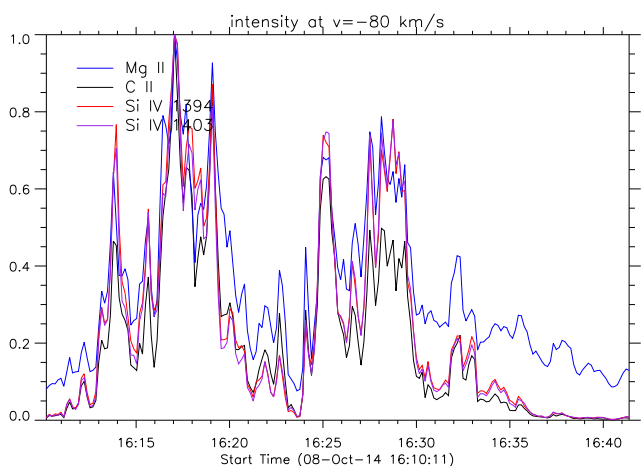}
   \caption{Example of (normalized) intensity evolution at $v=-80 $ km s$^{-1}$ of a burst showing that both the IB and the higher atmospheric burst occur simultaneously in all spectral lines. This is a typical behavior.}
        \label{lightcurve}
  \end{figure}

We inspected $>$30 bursts manually to better understand their temporal behavior, which is difficult to do in 
an automated way because of the large variety of available IRIS observational modes (rasters of different sizes and sit-and-stares), leading to very different cadences and thus often missing the details of the bursts' evolution.
Figure~\ref{fig6} shows two such examples. The wavelength is shown in Doppler units with the spectral lines indicated in the title of each panel. 
The time is on the y-axis and the intensities were root-scaled to increase their contrast. 
The left set shows that bursts can re-occur often within a few tens of seconds of one another, with the absorption lines remaining in place, except for small velocity shifts ($< 10$ km s$^{-1}$). 
The right example shows one IB with absorption line blends followed by a burst 20 minutes later in the same location but without any absorption blends (i.e., no cooler material on top). 
This second burst, and bursts of its type, were not included in the statistical analysis by our automatic selection. For all bursts we investigated, no \ion{Fe}{xxi} emission was noticeable, also not with any temporal delay. We only noticed an enhancement of the FUV continuum and of several FUV emission lines during bursts.

 \begin{figure*} 
  \centering 
   \includegraphics[width=.49\textwidth]{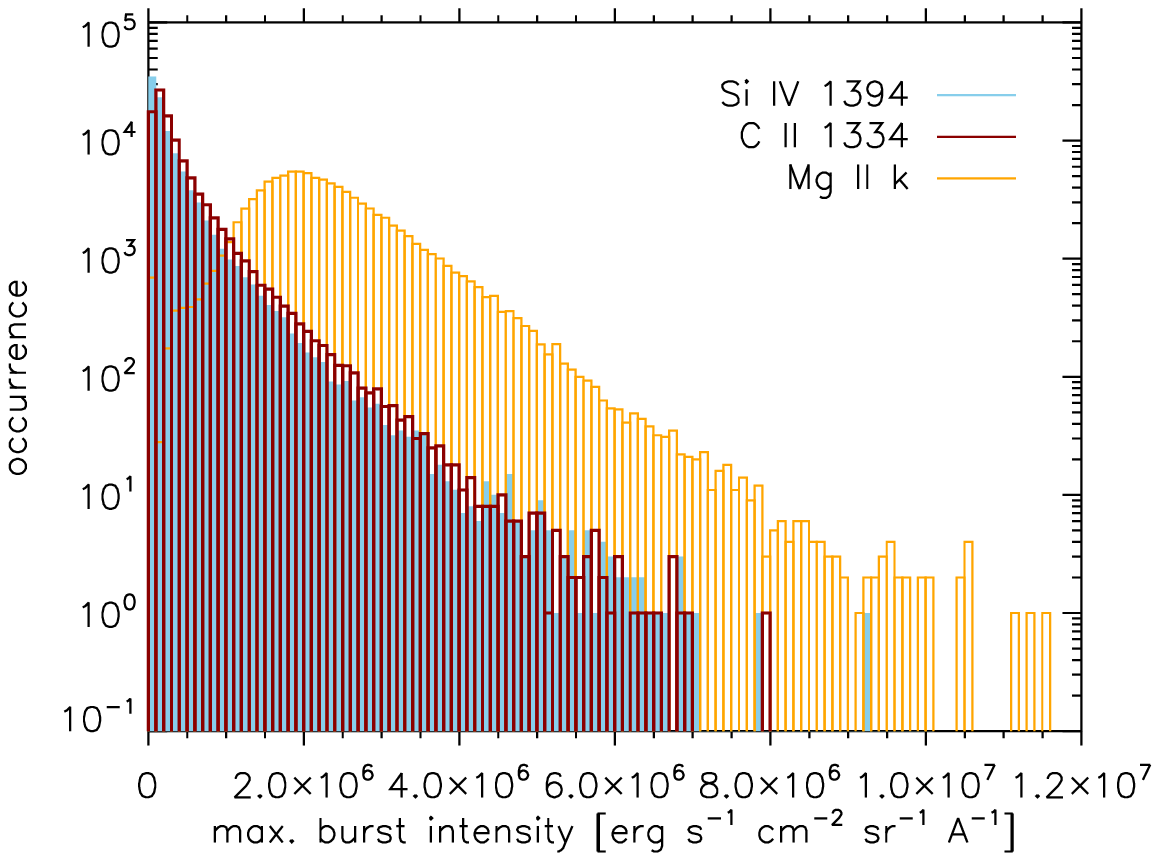}
   \includegraphics[width=.49\textwidth]{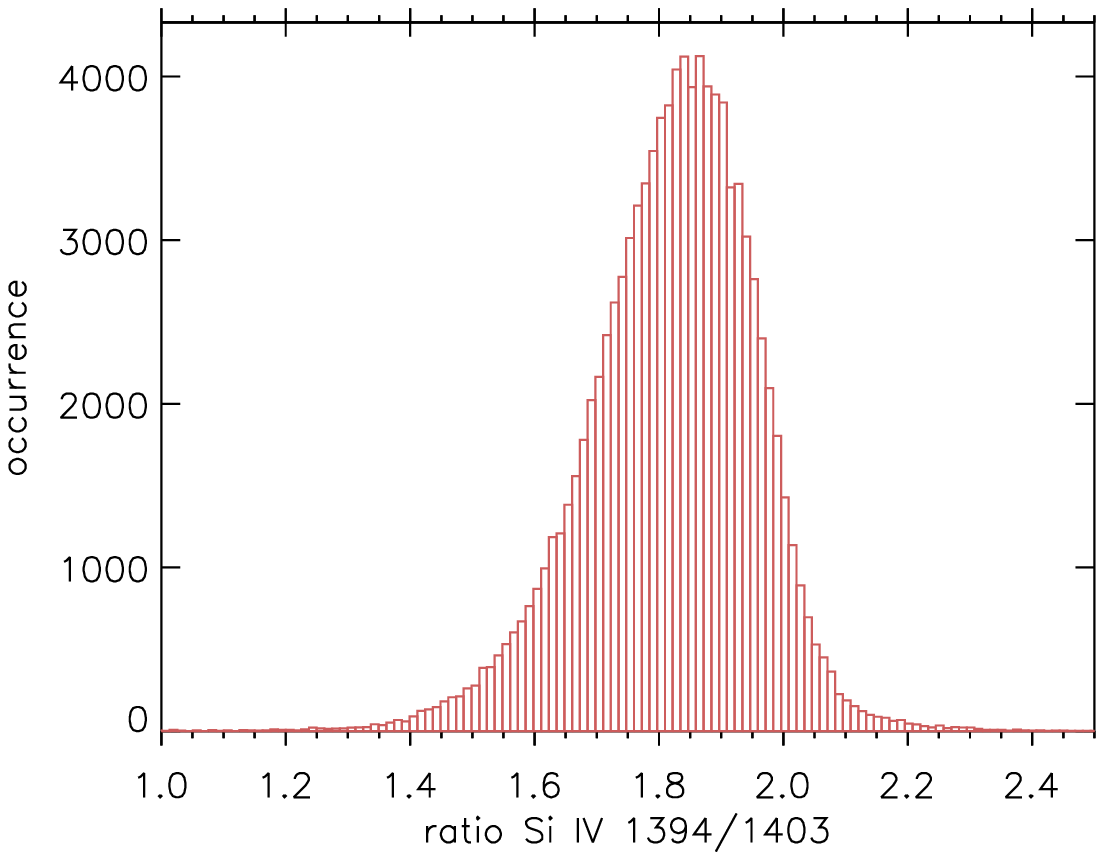}
   \caption{Left: Maximum intensities of bursts in cgs units. Right: \ion{Si}{iv} 1394 / \ion{Si}{iv} line ratio by integrating $\pm 1.5$\AA\ around each line core.}
        \label{maxint}
  \end{figure*}

 We also overplotted the lightcurves of bursts to investigate any potential time differences between the major spectral lines (\ion{Mg}{ii}, \ion{C}{ii} and both \ion{Si}{iv} lines), but there were no significant or systematic differences, an example of which is shown in Fig.~\ref{lightcurve}.

We converted the intensity into cgs units using the formulae described in \citet{kleintetal2016}. With this calibration, it is possible to analyze the burst intensities (maximum intensity of a spectral line) and line ratios via the integration of intensities at different wavelengths. The left panel of Figure~\ref{maxint} shows the maximum burst intensities. 
Since the distribution is continuous, we conclude that  selecting a cutoff energy for bursts may not be desirable. 
\ion{Si}{iv} 1394 and \ion{C}{ii} 1334 show similar burst intensities, while \ion{Mg}{ii} k is higher. 

We calculated the \ion{Si}{iv} 1394 / \ion{Si}{iv} 1403 ratio by integrating $\pm 1.5$ \AA\ around each line core. 
This method is not perfect because both lines have different absorption blends. 
We did however test the effect of the different blends on the calculated ratio by interpolating them away, and obtained only a $<3\%$ variation of line ratios.
The histogram of all \ion{Si}{iv} 1394 / \ion{Si}{iv} 1403 (without any blend removal) is shown in the right panel of Fig.~\ref{maxint}. 
The ratio peaks around 1.85, which is similar to the values reported by \citet{peteretal2014science} and \citet{yanetal2015}. 
A line ratio of 2 would be expected for optically thin conditions. However, optically thick conditions can lead to different values, from below 2 to above 2 \citep{tripathietal2020}. A more detailed analysis would be required for further conclusions based on line ratios.

\subsection{SDO/AIA wavelengths}\label{aia}

For each IB we extracted and analyzed 9 AIA lightcurves, averaged over an area of $5\times5$ AIA pixels, meaning $3\arcsec \times 3\arcsec$. All light curves for a given wavelength are of identical duration (12 or 24 minutes) and always centered at the AIA time step closest to the IB time. Because the spatial resolution of IRIS (0\farcs33) is higher than that of AIA ($\sim1.2\arcsec$), different IB spectra that occurred close to each other could result in identical AIA lightcurves, but this did not affect the analysis. 

 \begin{figure*} 
  \centering 
   \includegraphics[width=.85\textwidth]{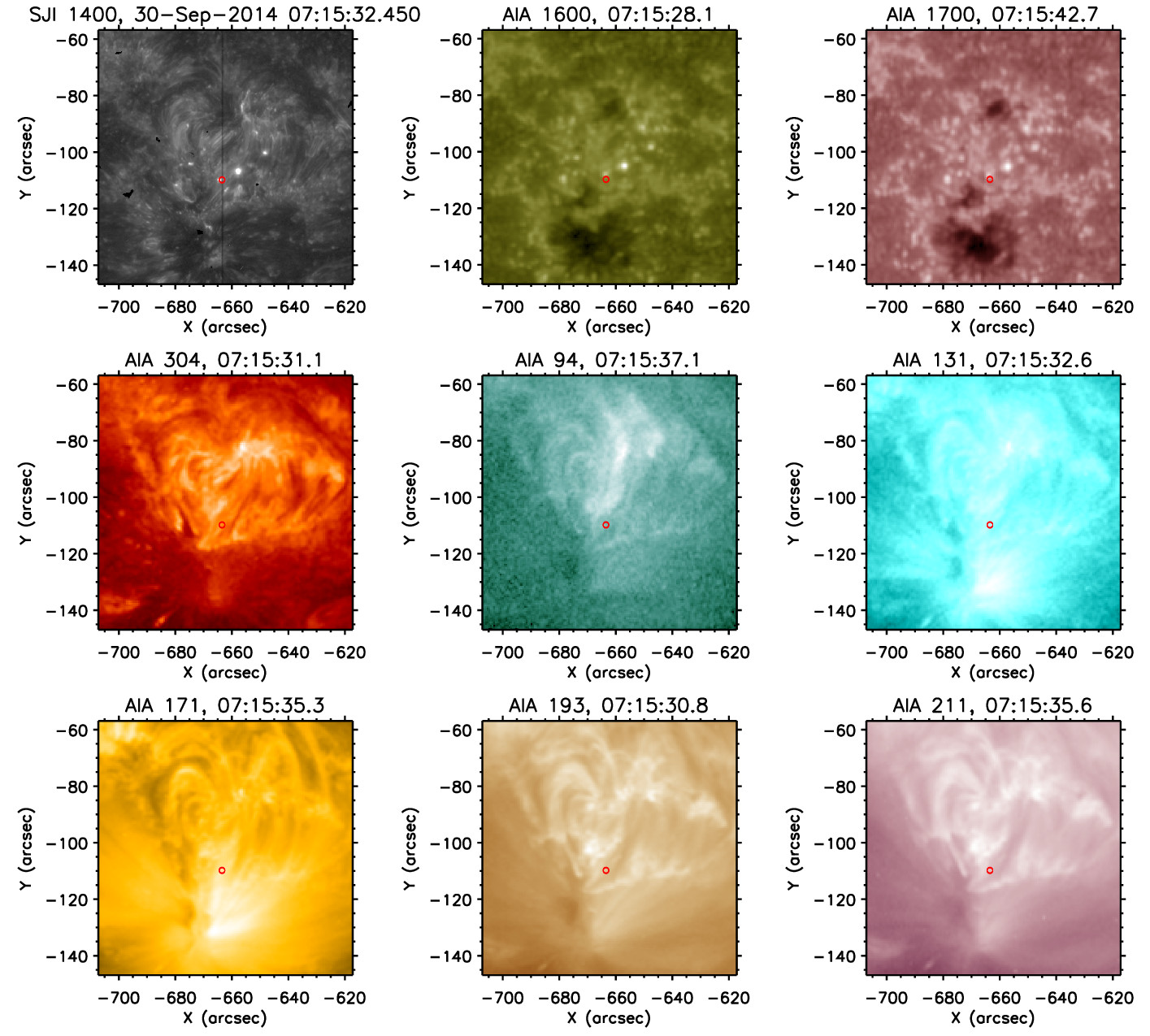}
   \caption{Example of locations of IBs (red circles). Co-spatial brightenings are visible in IRIS SJI, AIA 1600 and AIA 1700, while other AIA wavelengths do not seem to show any response to bursts. Because bursts were identified based on spectra, the red circles are always located close to the slit (vertical black line in SJI 1400). It is well possible that other brightenings in the SJI or AIA images also correspond to bursts, but that simply no spectra were available to identify them. An animation of this figure is available online.}
        \label{aiaoverv}
  \end{figure*}

\begin{figure*}[!htb]
\begin{center}
\includegraphics[width=0.82\textwidth]{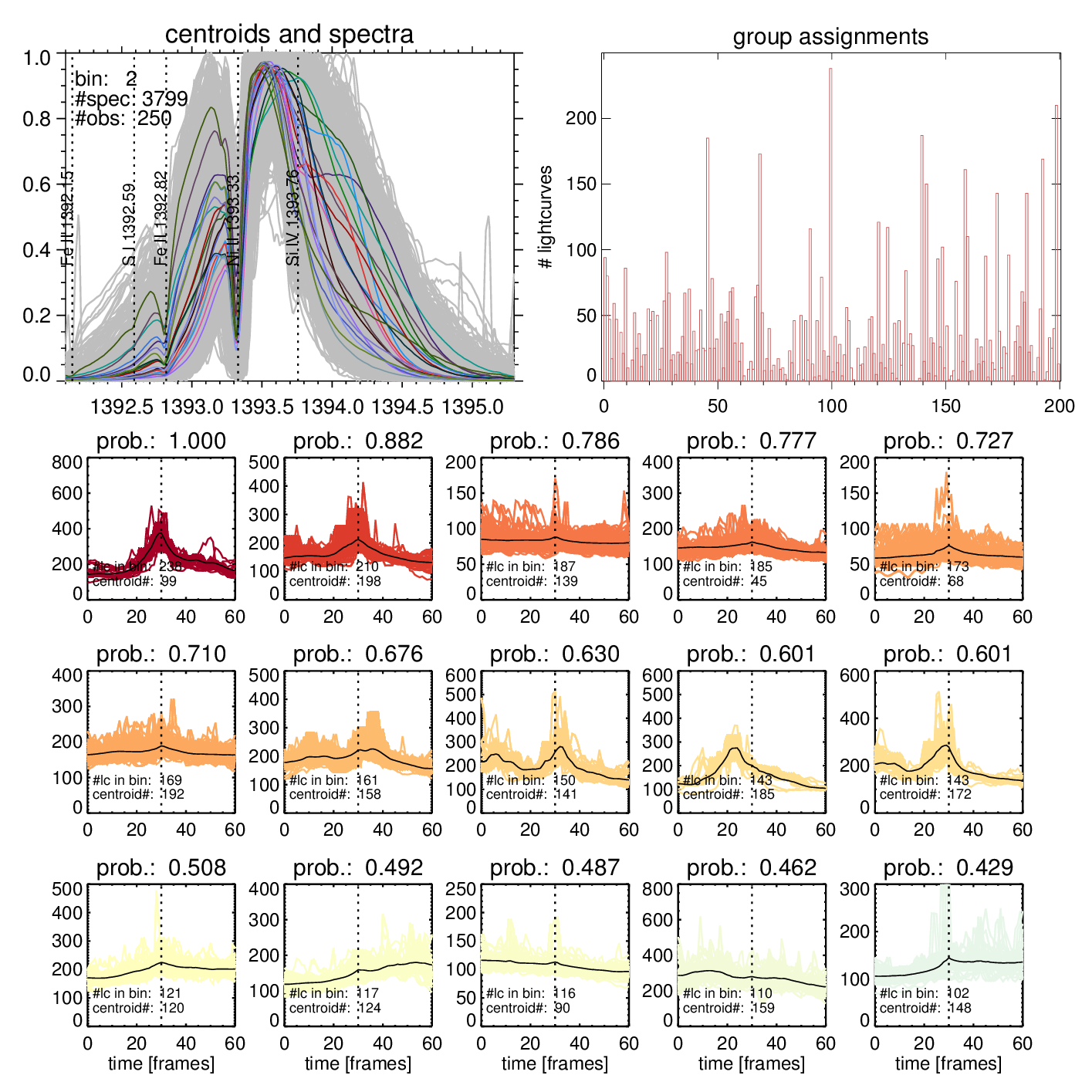}
\caption{Example of the light curve analysis, here for AIA 1600. For a given burst type (top left panel), the representation of the classification (k=200) is plotted in a histogram (top right panel). The panels below show 15 of the 200 light curve classifications, ordered by most- to least-occurring, with the same color-coding as used throughout the paper. In this example, it is obvious that most light curves show a peak at the 30th frame, which is marked by a vertical dotted line and corresponds to the IB time. This indicates that IBs are statistically observed simultaneously in IRIS and AIA 1600. Similar plots (not shown) were used for the analysis of all other AIA wavelengths and burst types.}\label{peaks}
\end{center}
\end{figure*}

We tested many different options how to best cluster the AIA lightcurves to obtain statistical results. 
Using k-means clustering on non-normalized AIA data, mostly similar intensities were grouped together, and small intensity increases due to bursts may be missed. 
This clustering gave us an overview of how the AIA intensities vary for IBs. 
On the other hand, using k-means clustering on normalized AIA data, results in a clustering that is more focused on the shapes of the lightcurves, meaning that e.g.\,enhancements that occur at a given time step are grouped together. The drawback of this version is that noisy lightcurves (lightcurves with random enhancements) are grouped into the real enhancement categories as well, especially for wavelengths such as AIA 94, which generally have very low intensities. 
For our analysis we therefore combined both the non-normalized and the normalized results. We were specifically searching for systematic enhancements in the lightcurves, either at or around the burst time $t_0$, as determined by IRIS spectra. A simultaneous enhancement in IRIS and AIA indicates co-existing temperatures, a later enhancement may indicate energy transfer to other atmospheric layers or heating. We did not exclude the possibility of AIA bursts starting before the IB time because, as IRIS rasters across the surface, the spectrograph slit may miss the beginning of a burst.
Figure~\ref{aiaoverv} shows an example of IBs that were identified. The red circles indicate their position. A circle is drawn at its $t_0 \pm 12$ seconds, which can be seen in the online animation of the figure, as bursts appear only close to the present location of the IRIS slit, which however does not mean that other brightenings further away from the slit in the images are not bursts. The figure and movie also show that AIA 1600 and AIA 1700 generally also show brightenings at burst locations, while the other AIA wavelengths do not seem to be correlated to bursts and also do not show delayed brightenings at burst locations. This behavior was in general also found via our statistical analysis, with a few exceptions.

For the analysis, we created plots as shown in Figure~\ref{peaks}, separately for all groups depicted in Fig.~\ref{fig2} (``burst types''), for all AIA wavelengths and for normalized and non-normalized lightcurves and analyzed them manually. The figure shows a non-normalized example (lightcurves in DN/s) for AIA 1600. The histogram on the top right shows the assignment of the lightcurves into the 200 different k-means groups. The 15 most populated groups (of the 200) are plotted on the bottom, color-coded by occurrences in the same color-table as Figs.~\ref{deepdip1} -- \ref{centerdip1}. It is clearly visible that most lightcurves show a peak at time step 30 (vertical dotted line), which corresponds to $t_0$. This means that bursts are statistically co-temporally visible in AIA 1600 and in IRIS spectra. In the case of AIA 1600, one time step corresponds to 24 s and the bursts in the light curves tend to last for about 10 time steps, meaning about 4 minutes. Our findings can be summarized as follows: (1) AIA 1600 generally shows enhancements at $t_0$ and while a few are weak or potentially even absent, there are no systematic differences between the burst types, (2) AIA 1700 often mirrors the behavior of AIA 1600, but the burst enhancements are weaker. In addition, burst types with bi-directional flows (e.g.~Fig.~\ref{centerdip1}) show a comparably minimal or absent response in AIA 1700. (3) all other AIA wavelengths do not show a statistically consistent behavior. A few percent of the light curves show peaks at $t_0$ in different AIA wavelengths and for different burst types, which most likely is random and/or noise. Only three burst types, marked by the blue frames in Fig.~\ref{fig2}, showed a significant number ($<50\%$ of the light curves) of peaks around $t_0$ in most AIA wavelengths, but even if these are true responses to IBs, they would only occur in fewer than 5\% of the overall light curves. In addition, the three burst types do not appear to have unique properties, except for relatively deep absorption lines, which however, also occur in other burst types where no significant AIA response was observed. In summary, while we cannot exclude that certain IBs may be visible in the AIA wavelengths 94, 131, 171, 193, 211, 304, 335, the vast majority is not. Since our analysis includes $t_0 \pm 6$ minutes for these wavelengths, we can also exclude a systematic delay (or advance) emission in these AIA wavelengths.


\section{Discussion}

\subsection{Occurrence and Classification}
This is the first large-scale statistical study of IB spectra. 
Our defining criterion of IRIS burst spectrum are absorption line blends, in particular the \ion{Ni}{ii} blend at 1393.33 \AA\ in the wing of \ion{Si}{iv} 1393.8 \AA, and we did not put any restrictions on the required intensities.
We found only $\sim 0.01\%$ of all IRIS spectra to be IRIS bursts, and thus they are relatively rare. Because of our large sample, our statistical conclusions are still based on more than 100000 IB spectra. We confirm previous findings of complex spectral shapes \citep[e.g.,][]{peteretal2014science, kimetal2015, tianetal2016} and add several new types of burst spectra. We find the blend absorption depth and the width of the \ion{Si}{iv} lines to vary significantly. There is no ``typical'' burst spectrum, rather the distribution between triangular, Gaussian, symmetric or asymmetric line shapes is relatively even. Similar spectral types usually appear in a large fraction of all burst observations, indicating that bursts are events with common physics, even if observed in different active regions and at different times and locations.

This study was able to determine co-occurring spectral shapes in other lines observed by IRIS (e.g.~in  \ion{C}{ii},  \ion{Mg}{ii} k and the triplet at 2791.60, 2798.75 and 2798.82 \AA, \ion{Fe}{ii} 2814.445 \AA) in a statistical way. The co-occurring spectral shapes vary as much as the different \ion{Si}{iv} burst types, but usually are correlated with each other. Meaning that if a downflow is observed in \ion{Si}{iv}, it is also likely to be visible in all other spectral lines. The one line that did not show any response to bursts was \ion{Fe}{xxi}, a hot coronal line \citep[e.g.][]{youngtianjaeggli2015}, indicating that IBs do not reach coronal temperatures.

\subsection{Temporal behavior}
The temporal behavior of the bursts is difficult to investigate statistically due to the wide range of observing modes, therefore our conclusions in this section are based on a smaller sample of bursts. Recurrent events were found regularly, as also  reported in past observations \citep[e.g.,][]{guptatripathi2015}.

Our sample also showed that bursts that are thought to form higher in the solar atmosphere, due to their lack of absorption blends, can occur in the same place as a previous IRIS burst. This may indicate either different magnetic topologies, either recurring reconnection in lower atmospheric layers, possibly due to flux emergence, or reconnection occurring higher up, either through existing or rising magnetic field lines.

We found the bursts to occur simultaneously in the spectral lines \ion{Mg}{ii}, \ion{C}{ii}, and \ion{Si}{iv}, which indicates that bursts are not only multi-thermal, but also that the heating to all temperatures occurs quite fast (seconds), otherwise hotter lines would be expected to show a delay.

\subsection{AIA lightcurves}
Correlated emission in AIA wavelengths, especially AIA 1600 and 1700 was found for EBs \citep[e.g.,][]{vissersetal2013} and for single observations of IBs \citep{guptatripathi2015}. The latter study even found emission in hotter AIA channels, but with a 5 minute delay and was unsure if it was correlated to the observed IB.

We confirm IBs to be  correlated with enhancements in AIA 1600 and AIA 1700 statistically. Other AIA wavelengths (94, 131, 171, 193, 211, 304, 335) do not show any consistent response to bursts, and in most cases, no response at all $\pm 6$ minutes from the burst time. It is therefore possible that the temporally delayed enhancement observed by \citet{guptatripathi2015} was either a very rare exception, or a coincidence. Our findings indicate that IBs do not show any significant response at coronal temperatures.

The enhancement of AIA1600 was found to be statistically higher than that of 1700. An explanation was suggested by \citet{youngetal2018} with the \ion{C}{iv} 1548 and 1500 \AA\ lines probably dominating compared to the continuum in the AIA 1600 filter passband for IBs.

\subsection{Comparison to models}
Various modeling efforts focused on burst spectra, mostly for EBs and some also for IBs \citep[e.g.,][]{reidetal2017, hongetal2017, hansteenetal2019}.  Simultaneous modeling of many spectral lines with different formation heights, optical thicknesses, and physical processes (e.g. complete redistribution CRD, or partial redistribution PRD) is very complex, and therefore not many studies exist to compare our burst spectra to. In general, to ionize silicon into the desired state, a temperature of around 25 kK is necessary in case of local thermal equilibrium, or 80 kK for lower densities \citep{rutten2016}. A thorough study was performed by  \citet{hongetal2021} who used a 1D radiative hydrodynamic simulation to investigate burst spectra by
 varying the height and shape of heating in their models. Their simulations of \ion{Si}{iv}, a Mg triplet line, and \ion{Mg}{ii} k allow us to directly compare our observations, at least qualitatively because the authors already concluded that the very broad \ion{Si}{iv} and \ion{Mg}{ii} spectra cannot be reproduced because of the limitations of 1D modeling. An interesting result is that the Mg triplet line is not affected for simulations of UV bursts, but only when the heating occurs lower in the atmosphere. In turn, low atmospheric heating does not cause sufficient \ion{Si}{iv} emission. Because our statistical analysis shows frequent enhancements in the  \ion{Mg}{ii} triplet lines and significant and broad \ion{Si}{iv} emission, this may indicate a wide height range of heating. The simulated  \ion{Mg}{ii} k spectra appear to have very deep core absorption compared to the observations, even though they were carried out with state-of-the-art simulations (PRD with the RH code). Such differences highlight the usefulness of statistical studies because they pose constraints on future modeling efforts.

\subsection{Potential biases}
Our analysis is based on all burst spectra and lightcurves and there is no extrapolation of any type of spectral shapes. The figures therefore show only real observed spectra. However, there may be potential biases in the way the data are grouped or displayed.

Our classification algorithm is based on the $L2$-norm. A cosmic ray in the line wing, or a large dip at line center may therefore lead to the same ``deviation'' from a reference spectrum and such spectra may be classified into a non-ideal category. By choosing our number of groups high enough, we reduce the in-class variance and by visually verifying the classification, we further checked the results. We cannot however guarantee that our IB list does not contain some false positives (non-burst spectra).

Another bias may be caused by our display of probabilities through the use of colors. The probabilities were assigned based on the frequency of co-occurring spectra grouped into certain k-means categories. If we imagine a fictitious example consisting of three k-means groups: a Gaussian-type spectrum and two downflow type spectra, one shifted by 20 km s$^{-1}$ and the other by 25 km s$^{-1}$. Then, due to the use of the $L2$-norm, the downflows, although they encode the same essential physics, will be classified into two distinct groups. Now let us suppose that 10 spectra are classified into each of these groups. In reality, we would like to see a conditional probability 1 for the downflows (20 spectra), to be drawn in red, and 0.5 for the Gaussian types (10 spectra), to be drawn in yellow. However, our method will result in a conditional probability of 1 for each of the groups (10 spectra each) and all would be colored in red. 
We verified our group assignments with histograms, visually, and with different binnings for the display to investigate the effect of such classifications and found them to be minor. Nevertheless, this is a bias to keep in mind when using unsupervised classification methods, since there is no guarantee that k-means groups are sufficiently different.


\section{Summary and Conclusions}

We analyzed more than 3500 IRIS observations to obtain statistics on IBs that were identified by absorption line blends on top of the \ion{Si}{iv} spectra, as first observed by \citet{peteretal2014science}. We found that bursts occur in $\sim$8\% of all analyzed observations, but only 0.01\% of all spectra observed by IRIS are bursts according to the above definition. Such bursts are thought to form in the lower solar atmosphere. We did not include any bursts without superimposed absorption lines in our analysis, which may form higher.

Our findings can be summarized as follows:
\begin{itemize}
\item We identified different types of burst spectra, which most likely result from small initial variation in burst conditions, such as: exact height, strength, geometry, or Doppler flows. A given burst spectral type could often be observed in hundreds of observations, hinting at a common underlying mechanism.
\item We found that other IRIS spectral lines also show peculiar spectra in bursts, often mirroring the behavior of the \ion{Si}{iv} line, e.g. deep absorption lines, strong downflow, etc. This indicates that IBs are multi-thermal in nature. There is no statistically significant response in \ion{Fe}{xxi}, thus restricting the IB temperature to below coronal temperatures. 
\item The temporal evolution shows recurrence of bursts, changes in their strength and likely height. \ion{Mg}{ii}, \ion{C}{ii}, and \ion{Si}{iv} show bursts simultaneously, indicating multi-thermal heating within seconds.
\item Emission in AIA 1600 and 1700 is correlated to bursts. Simultaneous emission in other AIA wavelengths is very rare, and not clearly attributable to bursts, at least within a timeframe of 12 minutes, further indicating a lack of coronal temperatures in IBs. 
\end{itemize}

Our statistical study shows the advantages of multi-wavelength studies in combination with automated algorithms. Our methods are easily transferable to any co-observed datasets and allow to derive statistical conclusions of a large amount of data. Our list of identified bursts is publicly available because further insights can be gained from this rich data set.

\begin{acknowledgements}
This work was supported by a SNSF PRIMA grant. We are grateful to LMSAL for allowing us to download the IRIS database and in particular to Ryan Timmons for his support. IRIS is a NASA small explorer mission developed and operated by LMSAL with mission operations executed at NASA Ames Research Center and major contributions to downlink communications funded by ESA and the Norwegian Space Centre. AIA is an instrument onboard NASA's Solar Dynamics Observatory. This research has made use of NASA's Astrophysics Data System Bibliographic Services.
\end{acknowledgements}

\bibliographystyle{apj}
\bibliography{journals,ibisflare}

\end{document}